\documentclass[submission, Phys]{SciPost}
\usepackage[utf8]{inputenc} 
\usepackage{XCharter}
\usepackage[T1]{fontenc}
\usepackage{amsmath,amssymb,amsfonts}
\usepackage{float}
\usepackage{color}
\usepackage[numbers,compress]{natbib}
\usepackage{graphicx}
\usepackage{caption}
\usepackage{mathrsfs}
\usepackage{multirow}
\usepackage{bm}
\usepackage{subcaption}
\usepackage[normalem]{ulem}
\newcommand{\magg}{\mathbf{M}}
\newcommand{\stagg}{\mathbf{N}}
\newcommand{\spin}{\mathbf{S}}
\newcommand{\Epseudo}{{\mathcal{E}}}

\newcommand{\magplus}{\mathcal{M}}

\newcommand{\invx}{\mathcal{X}}
\newcommand{\revertx}{\mathcal{I}}
\newcommand{\comnt}[1]{\ignorespaces} 
\newcommand{\dd}{\mathfrak{d}}
\newcommand{\half}{\frac{1}{2}}

\newcommand{\naturalset}{\mathbb{{N}}}
\newcommand{\comment}[1]{}
\hypersetup{
    colorlinks=true,
    linkcolor=blue,
    urlcolor=blue,
    citecolor=blue,
    anchorcolor=blue
}

\begin{document}


\begin{center}{\Large \textbf{
                Conservation laws and chaos propagation in a nonreciprocal classical magnet
	}}\end{center}
	
	\begin{center}
		Nisarg Bhatt\textsuperscript{$1$},
		Purnendu Das 
        \textsuperscript{$2,1$},
		Subroto Mukerjee\textsuperscript{$3$} and
        Sriram Ramaswamy\textsuperscript{$4,a$}
	\end{center}
	
	\begin{center}
		$^{1,3,4}$ Centre for Condensed Matter Theory, Department of Physics, Indian Institute of Science, Bengaluru 560 012, India 
		\\
		 and $^a$International Centre for Theoretical Sciences, Tata Institute of Fundamental Research, Bengaluru 560 089, India\\
         $^2$ Physics Department, TUM School of Natural Sciences, Technical University of Munich, 85748 Garching, Germany \\
$^1$ \href{mailto:bhatt@iisc.ac.in}{\small foundnisarg28@gmail.com},\quad
$^2$ \href{mailto:purnendu.das@tum.de}{\small purnendu.das@tum.de},\quad
$^3$ \href{mailto:smukerjee@iisc.ac.in}{\small smukerjee@iisc.ac.in},
$^4$\href{mailto:sriram@iisc.ac.in}{\small sriram@iisc.ac.in}
	\end{center}
    
\comnt{
\author{Nisarg Bhatt}
\email{bhatt@iisc.ac.in}
\author{Purnendu Das}
\email{purnendudas@iisc.ac.in}
\author{Subroto Mukerjee}
\email{smukerjee@iisc.ac.in}
\author{Sriram Ramaswamy$^*$}
\email{sriram@iisc.ac.in}
\affiliation{Centre for Condensed Matter Theory, Department of Physics, Indian Institute of Science, Bengaluru 560 012, India\\ and $^*$International Centre for Theoretical Sciences, Tata Institute of Fundamental Research, Bengaluru 560 089, India}
}


\begin{abstract}
{\bf
We study a nonreciprocal generalization [\href{https://iopscience.iop.org/article/10.1209/epl/i2002-00280-2}{EPL \textbf{60}, 418 (2002)}] of the classical Heisenberg spin chain, in which the exchange coupling is nonsymmetric, and establish numerically that it displays a 
ballistic spreading of chaos as measured by the decorrelator. We show that the interactions are reciprocal in terms of transformed variables, with conserved quantities that can be identified as magnetization and energy, with a Poisson-bracket algebra and Hamiltonian dynamics. For strictly antisymmetric couplings in the original model the conserved quantities diffuse, the decorrelator spreads symmetrically, and a simple hydrodynamic theory emerges. {When 
the interaction has symmetric and antisymmetric parts, 
the transformed description lacks translation invariance and a thermodynamic limit. Lastly, ballistic} propagation of chaos survives the inclusion of interactions beyond nearest neighbours, but the conservation laws in general do not. 
}
\end{abstract}
\tableofcontents
\section{Introduction} \label{sec:intro}
Effective interactions that evade Newton's 3rd Law are ubiquitous in living, active or driven matter \cite{ivlev2015statistical,dadhichi2020nonmutual, SSaha2020_nonreciCahnHilliard,You2020nonreciprocity,CahnHilliardnonvarcoupling2021,SSaha2019, LahiriSR1997, Uchida2010_microfluidicrotors, fruchart2021non,fruchart2026nonreciprocal}.
The importance of such nonreciprocity has been well understood in the context of learning and associative memory 
\cite{parisi1986asymmetric, Sompolinsky1986, Derrida1987, Rieger1988, Hennequin2012, Behera2023}, 
complex ecosystems \cite{Reichenbach2007, Allesina2012}, elastic networks \cite{Nassar2020, Scheibner2020, Zhou2020}, and non-Hermitian physics \cite{HatanoNelson96} 
The breaking of reciprocity leads to a host of highly original effects \cite{bowick2022symmetry, fruchart2026nonreciprocal,  avni2025nonreciprocal, PhysRevLett.117.248001} including  oscillations and travelling waves without mechanical inertia, 
\cite{simha1999traveling}, advection without a velocity \cite{dadhichi2020nonmutual,popli2025ordering}, pursuit-and-capture behaviour \cite{gupta2022active}, and exceptional-point phase transitions 
\cite{ 
PhysRevLett.122.185301, PhysRevResearch.2.033018, Galda2019,ramaswamy2002phase}. 

An early study \cite{das2002driven} explored criticality, chaos and control in the 
precessional dynamics {of 
classical} Heisenberg spins, with distinct exchange couplings for right and left neighbours {along one axis in space}. Although 
that work did not offer an exact derivation from a microscopic description, 
it argued that this manifestly nonreciprocal and non-Hamiltonian dynamics should arise generically in the presence of a sustained energy throughput, if
the underlying lattice was non-centrosymmetric {in the distinguished spatial direction}. 
Their work dealt almost entirely with a coarse-grained partial differential equation for the magnetization density, and not with the parent lattice model itself. 

The \textit{canonical} dynamics of classical Heisenberg spins \cite{ma1975critical} $\{\spin_n\}$ on sites $n$ of a lattice exhibits Larmor precession: $d\spin_n/dt = -\gamma \spin_n \times \mathbf{h}_n$, 
where $\gamma$ is the gyromagnetic ratio and the local magnetic field $\mathbf{h}_n = -\partial \mathcal{H} / \partial \spin_n$ arises from interactions with neighbouring spins through a Hamiltonian $\mathcal{H} = - \sum_{m<n}J_{mn}\spin_m \cdot \spin_n$. The reciprocal nature of this dynamics is clear: the torque on spin $n$ due to spin $m$ is precisely the negative of that on $m$ due to $n$. The model of \cite{das2002driven} amounts to retaining precessional dynamics, with $\mathbf{h}_n = \sum_m J_{mn} \spin_m$, but allowing a nonsymmetric $J_{mn}$ \footnote{Unlike the Dzyaloshinski\u{\i}-Moriya \cite{dzyaloshinsky1958thermodynamic,moriya1960anisotropic} coupling, this interaction does not mix space and spin.}. The dynamics still preserves spin-rotation symmetry but cannot be derived from the Hamiltonian $H$ above, which by construction can receive no contribution from the antisymmetric part of $J_{mn}$. In this paper we 
directly examine the dynamics of the lattice model of \cite{das2002driven}.
  
  
We study chaos propagation in our nonreciprocal model through the decorrelator, or Out-of-Time-Order Correlator (OTOC) \cite{larkin1969, maldacena2016bound, Huang2016, das2018light,  bilitewski2021classical, PhysRevX.8.031057}. 
We carry out extensive numerical studies on our nonreciprocally interacting spin models and find that the decorrelator spreads ballistically in every case.  
We will compare and contrast this behaviour with the chaos spreading with a butterfly velocity seen in studies of the classical one-dimensional nearest-neighbour Heisenberg model \cite{das2018light,McRoberts_PRB105}. 

Our main results are as follows: (i) The generic nonreciprocal model, with nearest-neighbour coupling, is symplectic and hence reciprocal when expressed in terms of a transformed spin variable. (ii) The model conserves the corresponding total magnetization of the transformed spin variables, and a corresponding energy function\footnote{Note that these conserved fields are distinct from quasi-conserved quantities as discovered in a class of periodically driven systems \cite{McRoberts_PRR5}.}. (iii) This escape from nonreciprocity comes at a price: the exponential position dependence of the change of variables thwarts a sensible thermodynamic limit for the general transformed description. Only for the purely antisymmetric model, which should therefore be viewed as an atypical special case, do the conserved quantities exist even under periodic boundary conditions and the passage to infinite system size pose no difficulty. (iv) We also extend the range of nonreciprocal interactions beyond the nearest neighbours and find that the conserved quantities in general cease to exist for such cases. (v) Chaos propagates ballistically in all the models studied, but the propagation speeds for the two directions are equal only for the antisymmetric model, {an observation} for which we provide an argument.

The rest of the paper is organized as follows: in section \ref{sec:NRdyn} we introduce the microscopic dynamics for the nonreciprocal models. We derive the conservation laws for the models with nearest neighbour couplings. In section  
\ref{sec:correlators} 
we present our numerical results on the two-point correlators of the densities of conserved quantities in the nonreciprocal models. In section 
\ref{sec:decorr} 
we present the numerical results on the decorrelator for our models, and the butterfly speeds and Lyapunov exponents obtained from them. 


\section{Nonreciprocal spin dynamics} \label{sec:NRdyn}
\subsection{Antisymmetric nearest-neighbour model} \label{subsec:antisymNN}
 We consider a system of classical Heisenberg spins ${\spin}_n$ of unit magnitude on a one-dimensional lattice {with sites labelled $n = 1,2,..., L$ where $L$ is even,} with periodic boundary conditions {unless otherwise specified}. With a purely antisymmetric exchange coupling between the neighbouring spins, $J_{n, n+1} = -J_{n+1,n} = 1$,  the dynamics is described by
\begin{align}
\dot{\spin_n} &= \spin_n \times (\spin_{n+1} - \spin_{n-1} )
\label{eqn:drvn_dynmk}
\end{align}
The dynamics obeys a Liouville-like theorem{, i.e., 
the flow in spin-space implied by \eqref{eqn:drvn_dynmk} is divergence-free; see \eqref{eqn:Liouville2_divergfree}}. However, it cannot 
be obtained from a Hamiltonian using the conventional Poisson bracket algebra for the spins ${\spin}_n$. It is straightforward to check that 
Eqn.(\ref{eqn:drvn_dynmk}) conserves neither the magnetization ${\magg}=\sum_r {\spin}_r$ nor the Heisenberg Hamiltonian $\mathcal{H} = -\sum_r \spin_r \cdot \spin_{r+1} $. Defining the staggered magnetization $\stagg_r \equiv (-1)^r {\spin}_r$, we see that the dynamics conserves $\stagg = \sum_r \stagg_r$ and  
the scalar 
$\mathcal{E}= \sum_r \mathcal{E}_r \equiv -\sum_r (-1)^{r} \spin_r \cdot \spin_{r+1}= \sum_r (-1)^{r} \stagg_r \cdot \stagg_{r+1}$, which we refer to as the pseudo-energy. Thus, quantities analogous to the conserved magnetization and energy in the Heisenberg model exist in our model. 

We show that the nonreciprocal dynamics in Eqn.\eqref{eqn:drvn_dynmk} can be expressed as {canonical dynamics 
by expressing the pseudo-energy in terms of the transformed variables and endowing the latter with a Poisson-bracket algebra. The resulting exchange coupling alternates spatially but is reciprocal, as it should be for dynamics governed by a Hamiltonian:} 
\begin{align}
\label{eqn:stag_poisson_hamil}
    \mathcal{H} &= \sum_r (-1)^{r} \stagg_r \cdot \stagg_{r+1}\, , \\
    \{N^{\alpha}_{r}, N^{\beta}_{n} \} &= \delta_{rn} \epsilon_{\alpha \beta \gamma} N_r^{\gamma}\, , 
\end{align}
\comnt{
\begin{align}
    \begin{split}
        \dfrac{\partial H}{\partial \stagg_n} &= (-1)^{r+1} (\delta_{r,n} \stagg_{r+1} + \delta_{r+1,n} \stagg_r ) \\
        &= (-1)^{n+1}\stagg_{n+1} + (-1)^n \stagg_{n-1} \\
        &= (-1)^{n+1} (\stagg_{n+1} - \stagg_{n-1})
    \end{split}
\end{align}
}
and therefore $\{S^{\alpha}_{r}, S^{\beta}_{n} \} \neq \delta_{rn} \epsilon_{\alpha \beta \gamma} S_r^{\gamma}$, {but rather $\{S^{\alpha}_{r}, S^{\beta}_{n} \} =(-1)^r\delta_{rn} \epsilon_{\alpha \beta \gamma} S_r^{\gamma}$}. Thus
\begin{align}
\label{eqn:drvn_staggdynmk}
    \begin{split}
    \dot{\stagg}_n &= \{\stagg _n, \mathcal{H}\} \\ 
    &= (-1)^{n+1} \stagg_n \times (\stagg_{n+1} - \stagg_{n-1})\,.
    \end{split}
\end{align}
It can be readily checked that the torques exerted by $\stagg_n, \stagg_{n\pm1}$ on each other in Eqn. \eqref{eqn:drvn_staggdynmk} are mutually opposite. Recalling that 
the total number of sites $L$ 
is even,  
we obtain the continuity equations for the two conserved quantities by 
{expressing the spins at 
alternate sites 
as linear combinations of direct and 
staggered magnetizations}: 
\begin{equation}
    \label{eqn:spin_MN_conversion}
    \spin_{2n}  = \magg_n + \stagg_n , \quad  \spin_{2n+1} = \magg_n - \stagg_n
\end{equation}


 \comnt{
\begin{align}
    \label{eqn:contnutMN_discrete} 
    \begin{split}
    \dot{\magg}_n(t) &=  \magg_n \times \half [(\dd_n \magg_n + \dd_n \magg_{n+1}) + \dd^2_n \stagg_n] - \stagg_n \times \half [(\dd_n \stagg_n + \dd_n \stagg_{n+1}) + \dd^2_n \magg_n] ,\\
    \dot{\stagg}_n(t) &= \stagg_n \times \half [(\dd_n \magg_n + \dd_n \magg_{n+1}) + \dd^2_n \stagg_n] - \magg_n \times \half [(\dd_n \stagg_n + \dd_n \stagg_{n+1}) + \dd^2_n \magg_n] , 
    \end{split}
\end{align}
where $\dd_n \magg_n = \magg_{n} - \magg_{n-1}$.  
}
{Applying the above transformation to the spin dynamics \eqref{eqn:drvn_dynmk} and coarse-graining, we obtain the following equations to lowest order in the gradients of $\magg$ and $\stagg$}:
\comnt{Coarse-graining the equations of motion upto first order in space derivatives 
yields  }
\begin{align}
    \label{eqn:contnutMN_continuum} 
    \begin{split}
        \partial_t{\magg}(x,t) &= \magg \times \partial_x \magg - \stagg \times \partial_x \stagg \\
        \partial_t{\stagg}(x,t) &= \stagg \times \partial_x \magg - \magg \times \partial_x \stagg  \\
        &= \partial_x (\stagg \times \magg) = -\partial_x 
        \bm{\mathcal{J}}^{\stagg}(x,t) \,. 
    \end{split}
\end{align}
The dynamics of the pseudo-energy density ${\mathcal{E}}_n = \frac{1}{2}\spin_{2n} \cdot (\spin_{2n-1} - \spin_{2n+1})$, can be similarly rewritten as 
\comnt{
\begin{equation} 
\label{eqn:Epseudo_MN_conversion}
    {\mathcal{E}}_n = \half\spin_{2n} \cdot (\spin_{2n-1} - \spin_{2n+1})
     = -\half(\magg_n+\stagg_n)\cdot \dd_n (\magg - \stagg)_n, 
\end{equation}
}
%
 \begin{align}
     \label{eqn:antisymm_Edyn}
    \partial_t \mathcal{E}(x,t) &= \partial_x [ (\magg \times \stagg) \cdot \partial_x (\magg + \stagg)] = -\partial_x \mathcal{J}^{\mathcal{E}}(x,t)
\end{align}

Eqns. \eqref{eqn:contnutMN_continuum} and \eqref{eqn:antisymm_Edyn} have the form of continuity equations, as expected, for the conserved quantities $\stagg$ and ${\mathcal{E}}$. {They are correct to first order in gradients, like the Euler equation for a fluid. However, they are not a purely hydrodynamic description as they contain the field $\magg$ which is not a slow variable.} 

\subsection{Nearest-neighbour nonreciprocal model} \label{subsec:NNNR} We present the nearest-neighbour nonreciprocal model whose dynamics is given by
\begin{align}
    \label{eqn:hybrid_dynmk1}
    \dot{ \spin}_n  &= \spin_n \times ( \spin_{n+1} + \alpha \spin_{n-1})
\end{align}
where $\alpha = \pm 1$ correspond to the Heisenberg and antisymmetric limits. 
We see that the linear combination $\boldsymbol{\mathcal{M}} = \sum_{n=1}^L \beta_n \spin_n$ 
is conserved if
\begin{align*}
\begin{split}
    \dot{\boldsymbol{\mathcal{M}}} &= \sum_n \beta_n \spin_n \times (\spin_{n+1} + \alpha \spin_{n-1}) = 0 \\
    &\implies \beta_n - \alpha \beta_{n+1} = 0 \quad, \forall n \\
    &\implies \beta_L = \alpha^{-L+1} \beta_1.
\end{split}
\end{align*}
Periodic boundary conditions would require $\alpha^L = 1$, which for {$\alpha$ real} 
{implies $\alpha = \pm 1$.}
For a general $\alpha$ we must work with open boundary conditions to preserve the conservation laws: $\dot{\spin}_1 = \spin_1 \times \spin_2, \quad \dot{\spin}_L = \alpha \spin_L \times \spin_{L-1}$, while the bulk obeys \eqref{eqn:hybrid_dynmk1}. 
We shall take $\beta_1 = 1$ for simplicity henceforth. The local conserved quantity $\boldsymbol{\mathcal{M}}_n = \alpha^{-n} \spin_n$ obeys a continuity equation in the bulk of the system, with its magnitude varying exponentially with $n$, in sharp contrast to the $\alpha = \pm 1$ cases.
The scalar $\mathcal{H} = \sum_r \mathcal{H}_r = -\sum_r \alpha^{-r} \spin_r \cdot \spin_{r+1} = -\sum_r \alpha^{r+1} \boldsymbol{\mathcal{M}}_r \cdot \boldsymbol{\mathcal{M}}_{r+1}$ is also conserved for this model, and can be viewed as an energy function as it generates a Hamiltonian dynamics with the spin Poisson-bracket algebra \eqref{eqn:poiss_stag} for the $\boldsymbol{\mathcal{M}}_n$. 
This can be checked easily:
\begin{align}
\label{eqn:poiss_stag}
    \{ \magplus^{\mu}_r, \magplus^{\nu}_n\} &= \epsilon_{\mu \nu \sigma} \delta_{rn} \magplus^{\sigma}_r
\end{align}
\begin{align}
    \label{eqn:hybrid2_magplus_variant1}   
    \begin{split}
    \dot{\boldsymbol{\magplus}}_n &= \{\boldsymbol{\magplus}_n, \mathcal{H} \} \\
    &= \alpha^{n+1} \boldsymbol{\magplus}_n \times \left( \boldsymbol{\magplus}_{n+1} + \dfrac{1}{\alpha} \boldsymbol{\magplus}_{n-1} \right)
    \end{split}
\end{align}
Note once again that $\{S^{\alpha}_{r}, S^{\beta}_{n} \} \neq \delta_{rn} \epsilon_{\alpha \beta \gamma} S_r^{\gamma}$.

\subsection{Further-neighbour interactions} \label{subsec:further} We now investigate the effect of 
interactions beyond nearest neighbours, for which we show that in certain cases 
conserved quantities of the sort discussed above do not exist. 
Consider the dynamics arising from antisymmetric coupling between each pair of spins under periodic boundary conditions:
\begin{equation}
\label{eqn:nnn_drvn_dynmk}
    \dot{\spin}_n = \sum_{r} J_r \spin_n \times (\spin_{n+r} - \spin_{n-r} )
\end{equation}
where $r \leq L/2 $.
A general linear combination of spins across all the sites can be written as
\begin{align}
    \boldsymbol{\sigma} &= \sum_n \gamma_n \spin_n,
    \label{eqn:sigma}
\end{align} 
where $\gamma_n$ is real. For conservation to hold, $\dot{\boldsymbol{\sigma}} = 0$. From the nearest neighbour case, we know that $\gamma_n = - \gamma_{n+1}$. Extending the spin interactions upto the next-nearest neighbour, we see that the $J_2$ term imposes  $\gamma_n = - \gamma_{n+2}$, which is not compatible with the first series $(\gamma_{n+2} = -\gamma_{n+1} = \gamma_n)$ to keep $\boldsymbol{\sigma}$ conserved. Thus, there is no conserved {magnetization-like quantity linear in the spins} for the dynamics governed by nearest- and next-nearest-neighbour antisymmetric interaction $(J_1, J_2 \neq 0)$. We see that upon setting the even-site interactions to zero $J_{2r} = 0$, while letting the odd-site interactions be non-zero $J_{2r-1} \neq 0$, $\boldsymbol{\sigma}$ is conserved once again: 
$\gamma_n = -\gamma_{n+1} = -\gamma_{n+3} = -\gamma_{n+5} \dots$. We will study the spin dynamics with next-nearest-neighbour terms in a later section in the context of chaos spreading.
The above results can be extended for the generic nonreciprocal interaction. We rewrite the dynamics for 
$\boldsymbol{\sigma}$, 
showing contributions from the symmetric and antisymmetric terms explicitly:
\begin{align} 
\label{eq:nnn}
\begin{split}
    \dot{\boldsymbol{\sigma}} 
    &= \sum_{n,r} \gamma_n \tilde{J}_r \spin_n \times (\spin_{n+r} + \alpha_r \spin_{n-r}),
    \end{split}
\end{align}
where  $\tilde{J}_r = J_r + J'_r, \text{ } \alpha_r = (J'_r - J_r)/(J'_r + J_r)$, with $(J_r)J'_r$ being the purely (non-)reciprocal coupling strengths. Conservation $(\dot{\boldsymbol{\sigma}} = 0)$ in \eqref{eq:nnn} requires $$\alpha_1 = \gamma_n / \gamma_{n+1}, \text{ } \alpha_2 = \gamma_n/\gamma_{n+2} = \alpha^2_1 , \text{ } \dots , \text{ } \alpha_r = \gamma_n / \gamma_{n+r} = \alpha_1^r, \, \dots  $$ 
Thus a conserved quantity linear in the spins exists for the generic nonreciprocal model if the above conditions are met. We now write a bilinear scalar, analogous to energy, as a linear combination of the local interaction terms:
\begin{align}
\label{eqn:eta}
\eta = \sum_{m,n} \zeta_{mn} \spin_n \cdot  \spin_m
\end{align}
where $\zeta_{mn} 
$ {are real}. Writing its time derivative
\begin{equation}
    \dot{\eta} = \sum_{r,n,m} J_r \zeta_{mn} \left[\spin_{m} \cdot \left(\spin_n \times (\spin_{n+r} - \spin_{n-r} )\right) + \spin_n \cdot \left( \spin_{m} \times (\spin_{m+r} - \spin_{m-r}) \right) \right] 
\end{equation}
which is in general non-zero. Thus, a bilinear scalar of the form \eqref{eqn:eta} is not conserved when the dynamics involves contributions from the nearest and the next-nearest neighbours $(J_1, J_2 \neq 0 )$ in an antisymmetric manner $(\alpha_1 = \alpha_2 = -1 )$. 

 

 
\section{Correlators of the conserved quantities} \label{sec:correlators} 
\subsection{Antisymmetric model} \label{subsec:antisym} 
\begin{figure*}[htp]
\centering
\begin{subfigure}{0.49\textwidth}
  \centering
  \includegraphics[width=\linewidth]{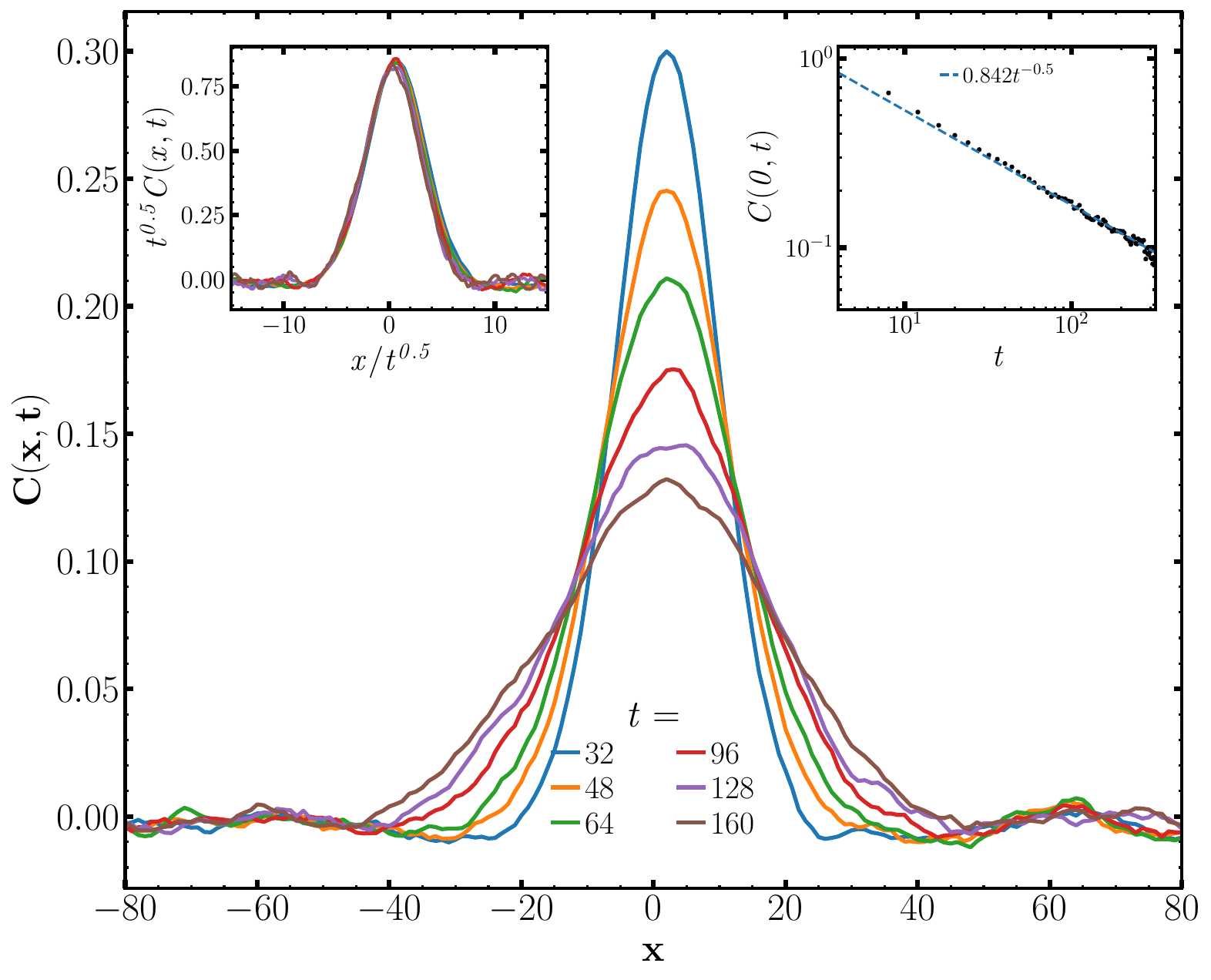}
  \caption{$C_{N}(x,t)$}
  \label{fig:Nxt_correlations}
\end{subfigure}\hfill
\begin{subfigure}{0.49\textwidth}
  \centering
  \includegraphics[width=\linewidth]{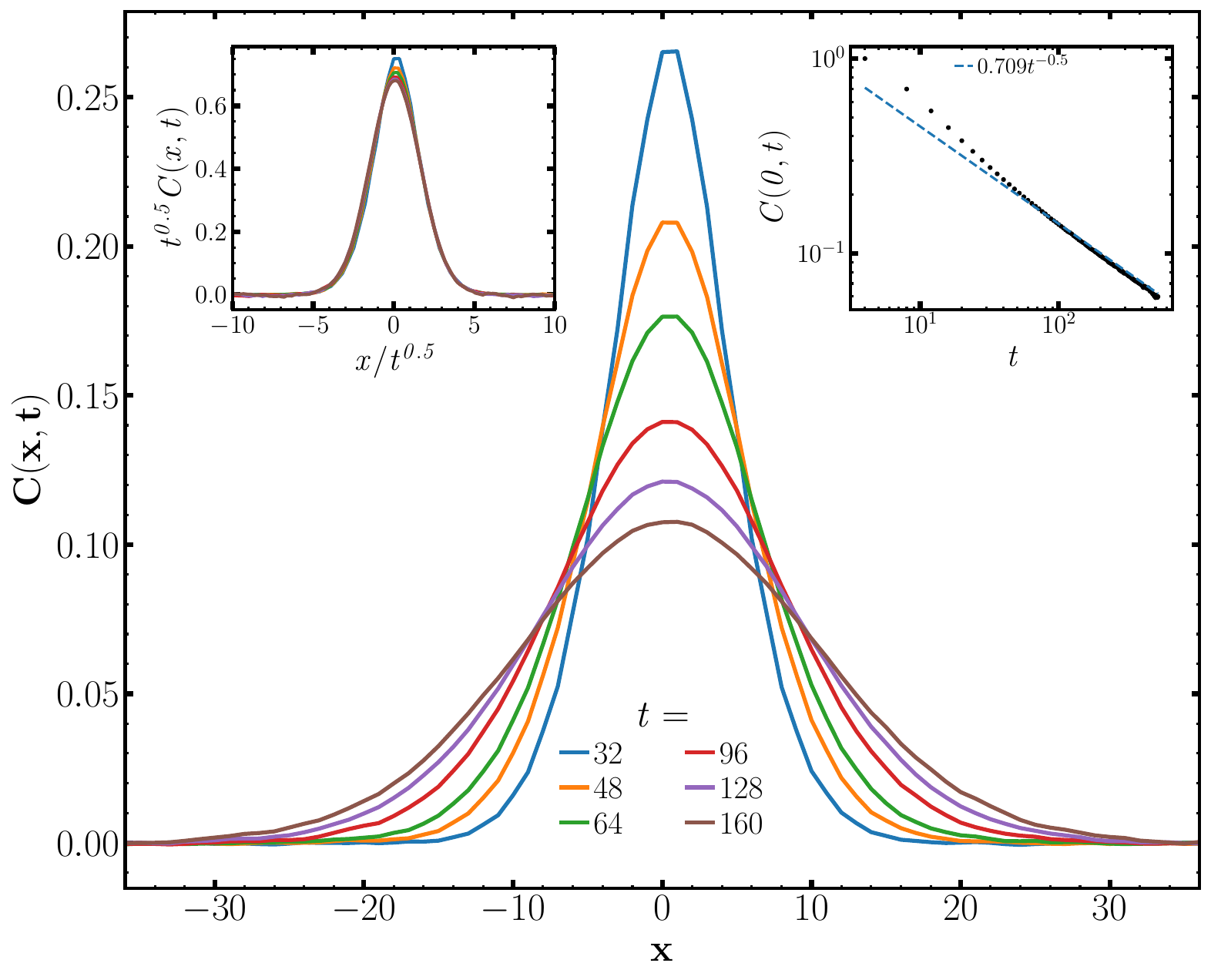}
  \caption{$C_{\mathcal{E}}(x,t)$}
  \label{fig:CENxt_correlations}
\end{subfigure}
\caption{\small Two-point correlation functions of (a) the staggered magnetization $C_{N}(x,t)$ and (b) the pseudo-energy $C_{\mathcal{E}}(x,t)$, with $x \in \{1,\dots, L\}$ for $L=512$, sampled over 5000 initial configurations as functions of $x$ for different values of $t$. The left inset of both panels shows the scaling collapse to a form $C(x,t) = t^{-1/2}f(x/t^{1/2})$ consistent with diffusion while the right inset shows a plot of $C(0,t)$ versus $t$ with a fit to $t^{-1/2}$.}
\label{fig:NN_EN_corrxt}
\end{figure*}

We now present our numerical calculation for the two-point correlators $C_N(x,t) = \langle {\stagg}(x+x_0,t+t_0) \cdot {\stagg}(x_0,t_0)\rangle$ and $C_{{\mathcal E}}(x,t) = \langle {\mathcal E}(x+x_0,t+t_0){\mathcal E} (x_0,t_0)\rangle$, evaluated in the statistical steady state, with 
samples 
drawn from an ensemble with equally weighted microstates.  
Each individual configuration of our system is evolved according to the dynamics of \eqref{eqn:drvn_dynmk} implemented through a RK4 integrator. The system size for the correlator calculation is $L= 512$, time-unit is $\Delta t = 0.001 - 0.002 $, evolved upto $320,000$ time-steps,  with  $5000$ initial conditions sampled  from uniformly distributed microstates. The correlator $C(x,t)$ is calculated by sliding over the reference point $x_0,t_0$ and averaging over 500 consecutive snapshots for each fixed time $t$, for each initial state.
The results are shown in Fig.(\ref{fig:NN_EN_corrxt}), where the 
data collapse suggests a scaling consistent with diffusion for the correlations of ${\stagg}$ and of ${\mathcal E}$. 

%
The two-point correlator of the conserved quantities exhibits a scaling form $C(x,t) = t^{-\nu} f(t^{-\nu}x)$, with scaling exponent $\nu$ and scaling function $f$ which we expect characterizes a universality class of systems, not only the particular microscopic model we have studied. 
The numerical results for $C_{N}(x,t)$ and $C_{\mathcal{E}}(x,t)$ in Fig.(\ref{fig:NN_EN_corrxt}) show that the two-point correlators for both the conserved quantities decay in time in a manner consistent with diffusion ($\nu = 1/2$). The insets show a scaling collapse of both correlators to the diffusive form $t^{-1/2} f(t^{-1/2}x)$, with different functions $f$ for the two cases. 
As in the classical Heisenberg chain, the presence of a finite number of conserved quantities for this model points to the non-integrability of the system, which is consistent with the observation of diffusion. {Analytical arguments establishing 
the diffusive 
character of the conserved fields} 
and the transient nature of the fast modes 
within a hydrodynamic framework will be reported in a separate work. 
A numerical result showing the rapid decay of the magnetization {$\sum_n {\bf S}_n$} in the antisymmetric exchange model, starting from {a range of} 
initial states is shown in Fig.(\ref{fig:Mkt_decayvs_t}) of  Appendix \ref{App:continuity}. 

\subsection{The generic nonreciprocal model} \label{subsec:generic_NR}
\begin{figure*}[htp]
\centering
\begin{subfigure}{0.49\textwidth}
  \centering
  \includegraphics[width=\linewidth]{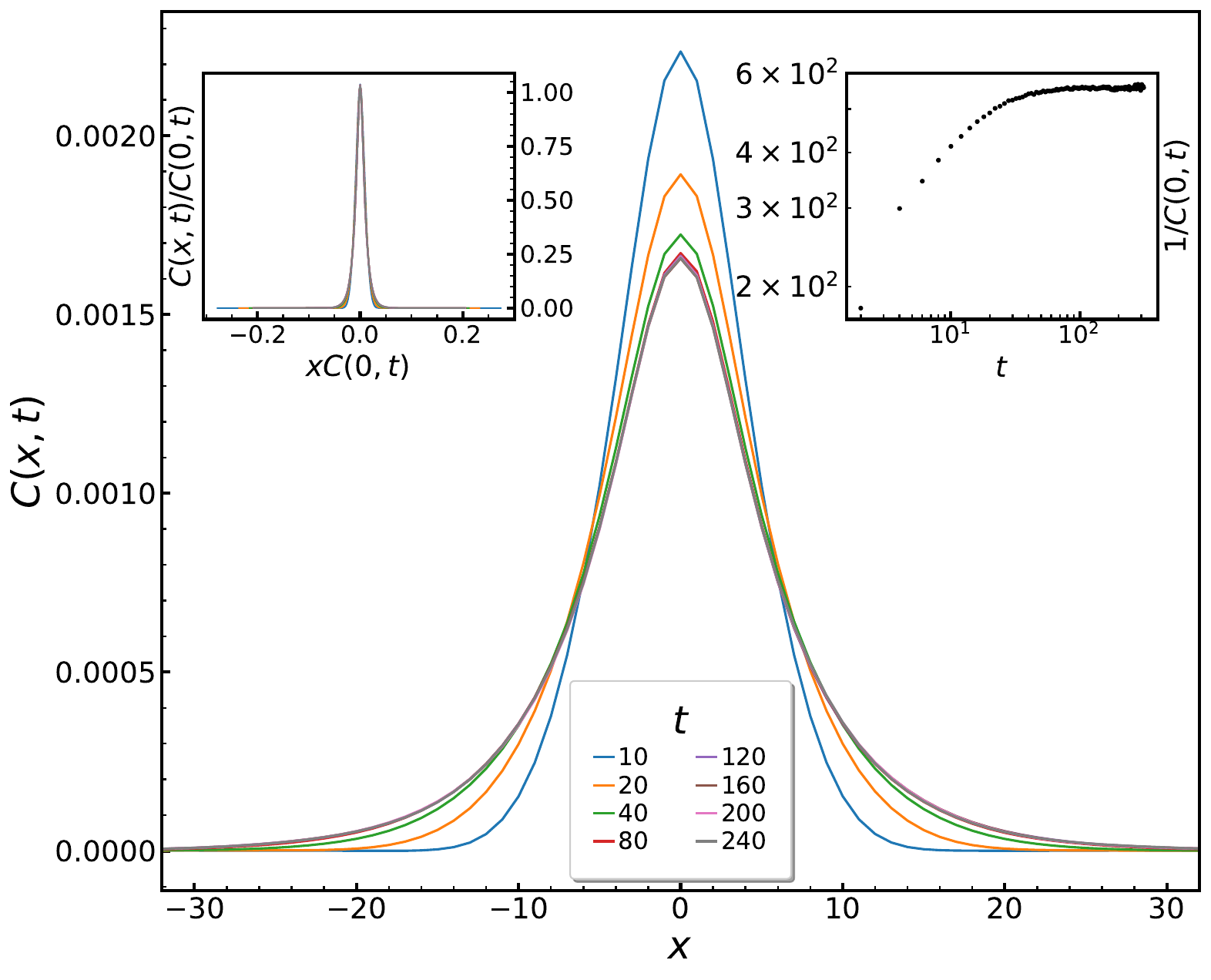}
  \caption{$C_{\magplus}(x,t)$}
  \label{fig:Magplus_correlations_1}
\end{subfigure}\hfill
\begin{subfigure}{0.49\textwidth}
  \centering
  \includegraphics[width=\linewidth]{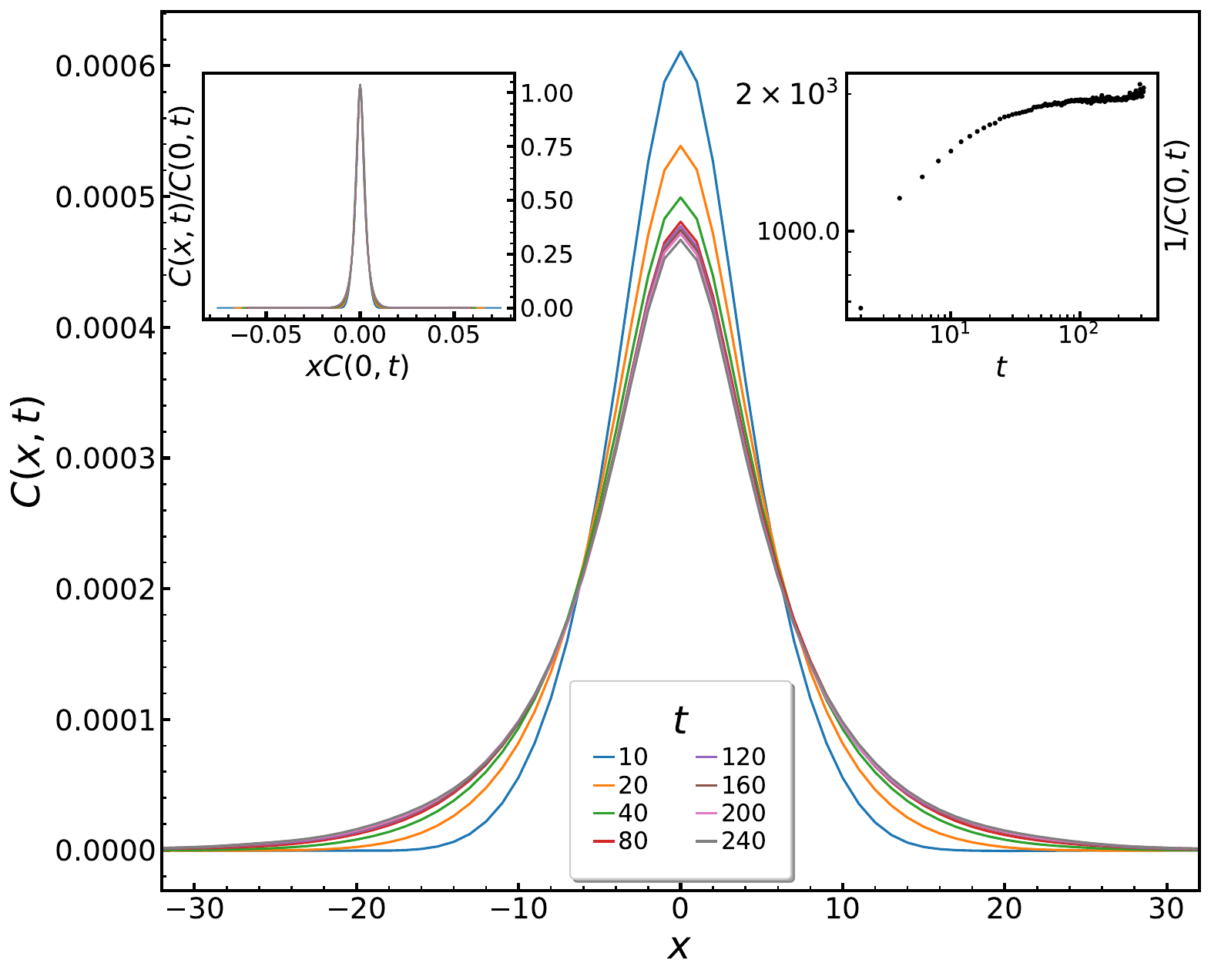}
  \caption{$C_{\mathcal{E}}(x,t)$}
  \label{fig:Epseudo_correlations_1}
\end{subfigure}
\caption{ \small Two-point correlation functions: (a) $C_{\magplus}(x,t)$ and (b) $C_{\mathcal{E}}(x,t)$ for $\alpha=1.1$, with $x \in \{1,\dots, L\}$, $L=256, \text{ and} \Delta t = 0.002$, sampled over 5000 initial configurations as functions of $x$ for different values of $t$. There 
is no power law that scales $C(x,t)$ at different time steps{, 
as seen in the saturation of 
$C^{-1}(0,t)$ vs $t$ (right insets). Taking $C^{-1}(0,t)$ as the length scale, we see that the plot of 
$C(x,t)/C(0,t)$ against $x C(0,t)$ 
shows a collapse.}}
\label{fig:MM_EM_corrxt_1}
\end{figure*}

\begin{figure*}[htp]
\centering
\begin{subfigure}{0.49\textwidth}
  \centering
  \includegraphics[width=\linewidth]{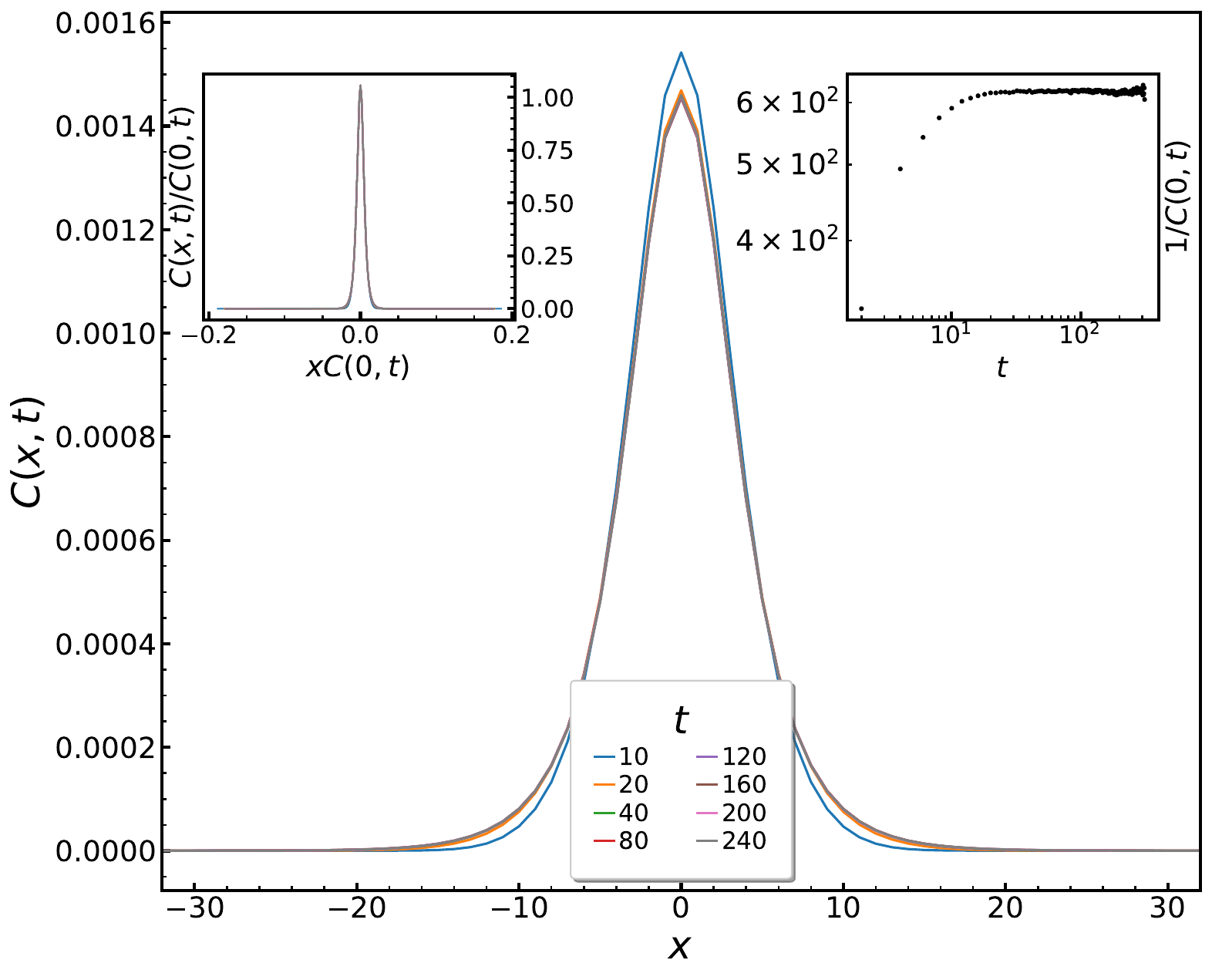}
  \caption{$C_{\magplus}(x,t)$}
  \label{fig:Magplus_correlations_2}
\end{subfigure}\hfill
\begin{subfigure}{0.49\textwidth}
  \centering
  \includegraphics[width=\linewidth]{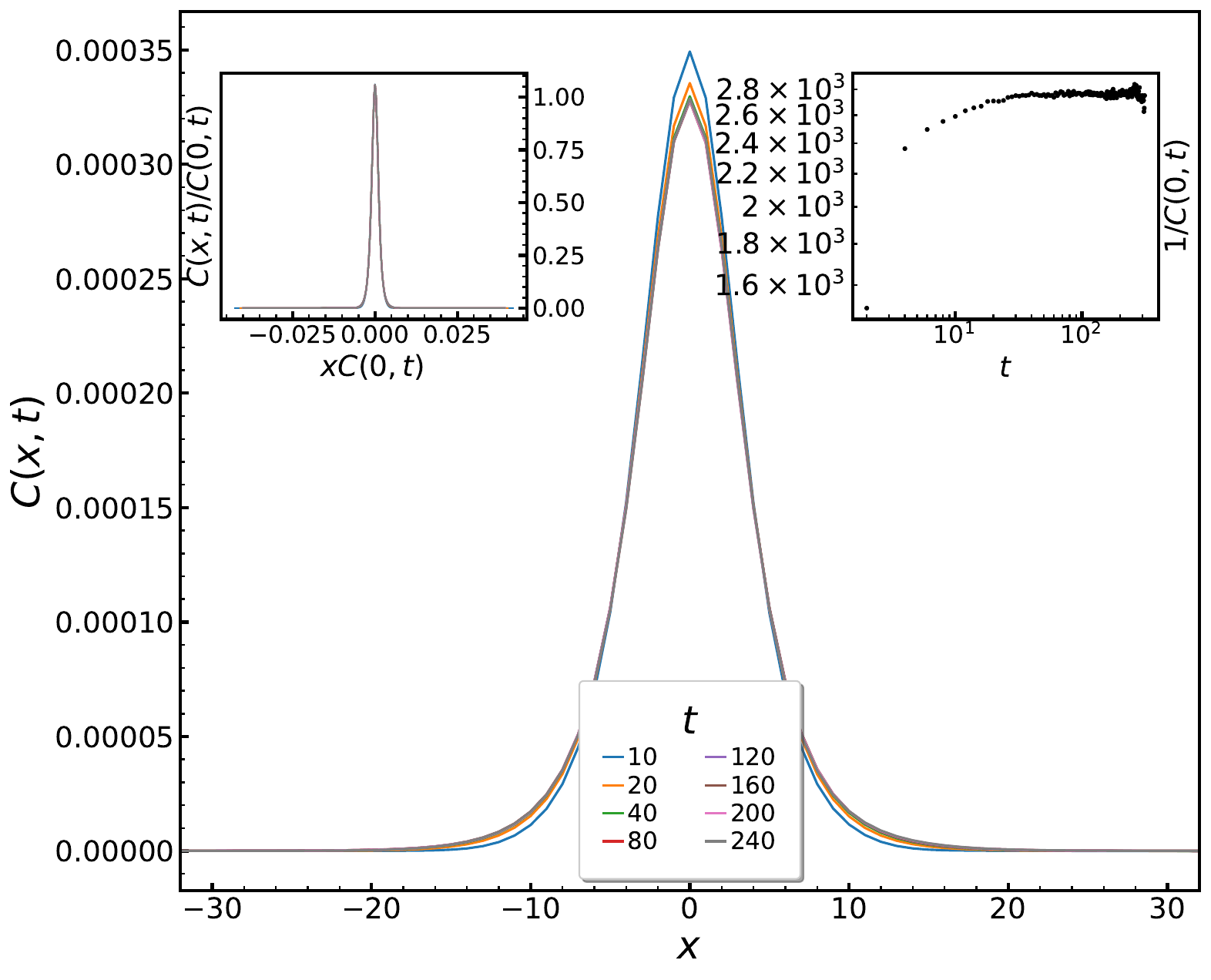}
  \caption{$C_{\mathcal{E}}(x,t)$}
  \label{fig:Epseudo_correlations_2}
\end{subfigure}
\caption{ \small Two-point correlation functions of: (a) the conserved magnetization, $C_{\magplus}(x,t)$ for $\alpha=1.2$ and (b) the pseudo-energy, $C_{\mathcal{E}}(x,t)$, with $x \in \{1,\dots, L\}$ for $L=256$, sampled over 5000 initial configurations as functions of $x$ for different values of $t$. }
\label{fig:MM_EM_corrxt_2}
\end{figure*}

We now numerically determine the correlator for the conserved quantities for the generic nonreciprocal model described by \eqref{eqn:hybrid_dynmk1}. The numerical results are studied under open boundary conditions, and are carried out in the vicinity of $|\alpha| = 1$,  since the exponential $x$-dependence of the coefficients $\alpha^{-x}$ limits the numerical precision of the calculation. System size is taken to be $L=256$. {We take $|\alpha| > 1$ without loss of generality since $\alpha \rightarrow 1/\alpha $ simply corresponds to switching the strengths of the left- and right-neighbour interactions at each site}. The correlation function for the density $\magplus_n =  \alpha^{-n}\spin_n $ of the conserved quantity linear in $\spin$ is given by
\begin{equation}
    \begin{split}
    C_{\magplus}(x,t) = \left \langle \mathop{\sum\nolimits'}_n 
    \magplus_{n+x}(t) \magplus_n(0)
    \right \rangle
    \end{split}
    \label{equ:corr_generalized_linearS}
\end{equation}
where the $\sum_n'$ denotes that the range of summation depends on the value of $x$. The site indices are $ n = 1,2,\dots, L$. For $x\geq 0$, $\sum_n' = \sum_{n = 1}^{L-x}$ and for $x<0$, $\sum_n' = \sum_{n = -x+1}^{L}$.
The spread of the correlator $C_{\magplus}(x,t)$ with time is shown in Figures \ref{fig:MM_EM_corrxt_1}(a), \ref{fig:MM_EM_corrxt_2}(a). As expected, the area under the curve, which gives the quantity $A_{\magplus}  = \sum_{x = -L+1}^{L-1} C_{\magplus}(x,t)$, remains constant with time. One can see that by simply expanding the expression of $A_\mathcal{M}(t) =  \sum_{x = -(L-1)}^{L-1} \left \langle \mathop{\sum\nolimits'}_n\frac{1}{\alpha^{n+x}}\spin_{n+x} (t) \cdot \frac{1}{\alpha^{n}}\spin_n (0) \right \rangle =\left \langle \left(\sum_{x=0}^{L-1}\frac{1}{\alpha^x}\spin_x (t)\right)\left(\sum_{n=0}^{L-1}\frac{1}{\alpha^n}\spin_n (0)\right)\right \rangle$. The first term in the parentheses represents the conserved quantity which is linear in $\textbf{S}$. It is the only term with a time index, and therefore its time derivative is zero. 

We also calculate the correlation function 
\begin{equation}
\label{pseudoenergycorr}
    \begin{split}
    C_{\mathcal{E}}(x,t) = \left\langle \mathop{\sum\nolimits''}_n \mathcal{E}_{n+x}(t)\mathcal{E}_{n}(0)\right\rangle = \left\langle \mathop{\sum\nolimits''}_n\frac{1}{\alpha^{n+x+1}}\spin_{n+x} (t) \cdot \spin_{n+x+1} (t)\frac{1}{\alpha^{n+1}}\spin_{n} (0) \cdot \spin_{n+1} (0)\right\rangle 
    \end{split}
\end{equation} 
for the density $\mathcal{E}_n(t) = \frac{1}{\alpha^{n+1}} \spin_n(t)\cdot \spin_{n+1}(t)$ of the conserved quantity quadratic in $\spin$.
The $\sum_n''$ denotes that the summation range depends on the value of $x$ as follows. For $x>0$, $\mathop{\sum\nolimits''}_n = \sum_{n = 1}^{L-x-1}$ and for $x<0$, $\mathop{\sum\nolimits''}_n = \sum_{n = -x+1}^{L-1}$
The spread of the correlator $C_{\mathcal{E}}(x,t)$ with time is shown in Figures [\ref{fig:MM_EM_corrxt_1}(b)], [\ref{fig:MM_EM_corrxt_2}(b)]. As $\mathcal{E}$ is a conserved quantity, the area under the curve, $A_{\mathcal{E}}  = \sum_{x = -(L-2)}^{L-2} C_{\mathcal{E}}(x,t)$, is also conserved. This can be proven in a manner similar to $A_{\mathcal{M}}$ \eqref{eqn:ConservedArea_AE}.

Unlike for the Heisenberg and antisymmetric exchange models, there is no sustained spreading of the correlator here, as visible from the stagnation of the peaks of the correlators at later times. The correlator peak as a function of time corresponds to an inverse length scale by which the two-point correlators of the conserved quantities at different times can be scaled to yield a collapse. The absence of a power-law growth of the length scale is seen in the saturation in the inverse peak values, pointing to an inherent length scale in the system which impedes diffusion of the conserved quantities. From Eqn. \eqref{eqn:hybrid2_magplus_variant1} we know that the coupling constant in the dynamics in terms of the transformed variables is site dependent and thus provides a length-scale proportional to $1/|\ln|\alpha||$.

To highlight another interesting feature of the generic nonreciprocal model, we plot the site-anchored correlation functions among the components $\magplus_n^{\alpha}$, i.e. 
\begin{align}
\label{eqn:corr_site_magplus}
    C^{\magplus}_n(x,t) &= \left \langle \frac{1}{\alpha^{n+x}}\spin_{n+x} (t) \cdot \frac{1}{\alpha^{n}}\spin_n (0) \right \rangle
\end{align}
\textit{i.e., } without performing the sum over lattice sites $n$. This site-anchored correlator depends on the site index $n$, and $x$ can be any integer between $-n+1$ and $L-n$. $C^{\magplus}_n(x,t)$ exhibits directionality or drift in its spatial spread, i.e., the spread is no longer symmetric about $x=0$ (see Figure[\ref{fig:MM_corrxt_3_unconserved}]). 
However, the spatial spread of the 
site-averaged correlator $C_{\magplus}(x,t)$ is symmetric about $x =0$. This symmetry arises because the site-anchored correlator $C^{\magplus}_n(x,t)$ of the larger site index $n$ contributes more towards the negative $x$ values. When summed over all $n$, these asymmetric individual contributions collectively yield a total correlator $C_{\magplus}(x,t)$ that is symmetric about $x = 0$. To 
show analytically that $C_{\magplus}(x,t)$ is spatially symmetric, we first prove that it is time-reversal symmetric, i.e. $C_{\magplus}(x,t) = C_{\magplus}(x,-t)$. If we invert the time i.e. $t \to -t$ and also make the substitution ${\tilde{\spin}_n(t)} = -\spin_n(-t)$, then it will leave the form of Eqn.\eqref{eqn:hybrid_dynmk1} unchanged. Thus, one can directly write $C_{\magplus}(x,t)  =\langle \mathop{\sum\nolimits'}_n\frac{1}{\alpha^{n+x}}\bm{\tilde{S}}_{n+x} (t) \cdot \frac{1}{\alpha^{n}}{\tilde{\spin}}_n (0)\rangle
        =\langle \mathop{\sum\nolimits'}_n\frac{1}{\alpha^{n+x}}(-\spin_{n+x} (-t)) \cdot \frac{1}{\alpha^{n}}(-\spin_n (0))\rangle
          = C_{\magplus}(x,-t)$
Hence, the correlator is time-reversal symmetric.

Next, we will show $C_{\magplus}(x,t) = C_{\magplus}(-x,-t)$.
Without loss of generality, consider $x> 0$. Then, using time translational symmetry, one can write 
    $C_{\magplus}(x,t) = \langle \sum_{n=1}^{L-x}\frac{1}{\alpha^{n+x}}\spin_{n+x} (0) \cdot \frac{1}{\alpha^{n}}\spin_n (-t)\rangle
    = \langle \sum_{n'=x+1}^{L}\frac{1}{\alpha^{n'}}\spin_{n'} (0) \cdot \frac{1}{\alpha^{n'-x}}\spin_{n'-x}(-t)\rangle 
     = C_{\magplus}(-x,-t)$
Hence, we have shown that site-averaged correlator is symmetric about $x=0$
\begin{equation}
\label{eqn:summedcorr_symm}
C_{\magplus}(x,t) = C_{\magplus}(x,-t) = C_{\magplus}(-x,-t) 
\end{equation}
A more detailed calculation is shown in Appendix \ref{App:symm_correlator}.

\begin{figure*}[htp]
\centering
\begin{subfigure}{0.45\textwidth}
  \centering
  \includegraphics[width=\linewidth]{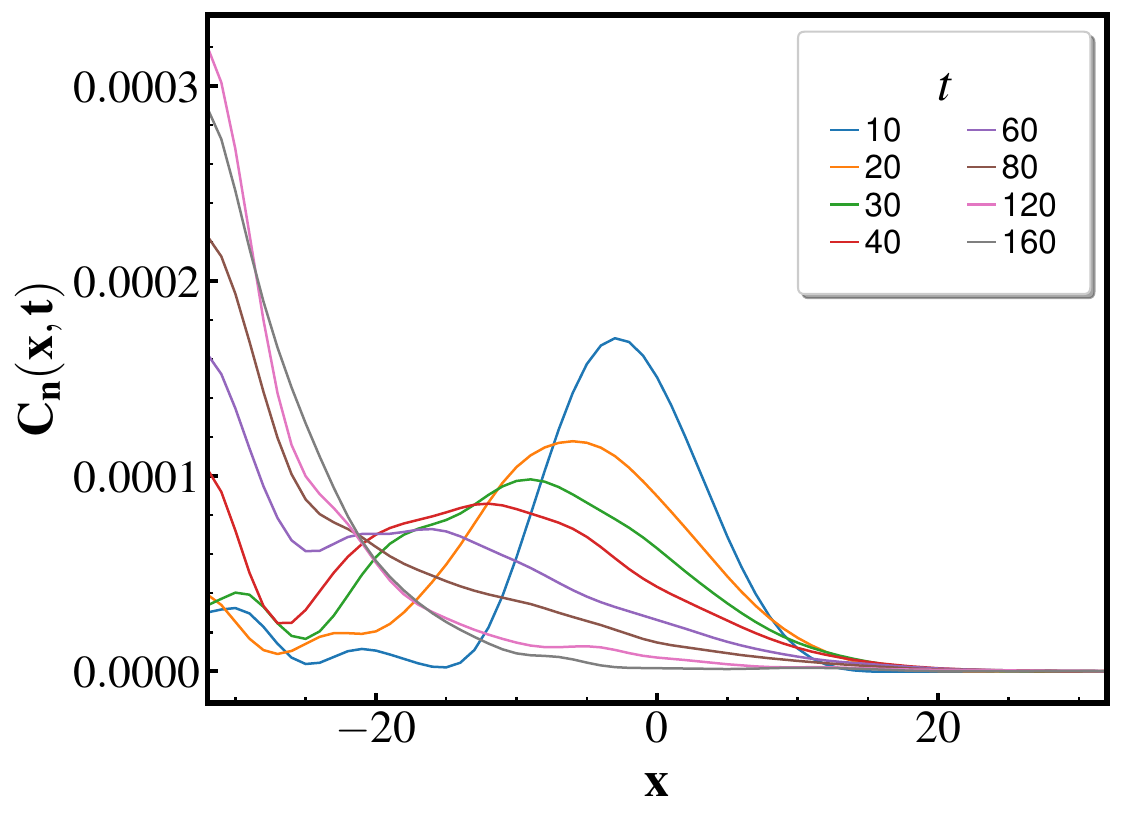}
  \caption{}
  \label{fig:CN_n_L8}
\end{subfigure}\hfill
\begin{subfigure}{0.45\textwidth}
  \centering
  \includegraphics[width=\linewidth]{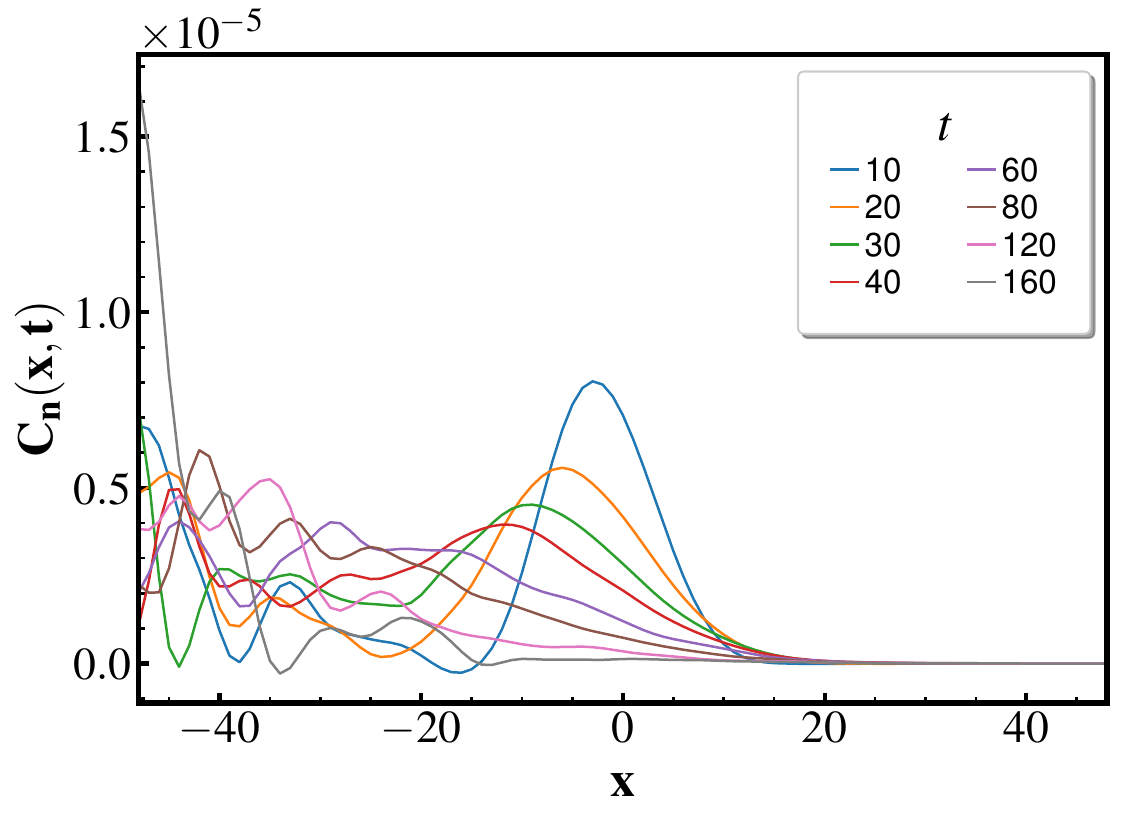}
  \caption{}
  \label{fig:CN_n_3L16}
\end{subfigure}
\caption{\small Site-anchored correlator $ C^{\magplus}_n(x,t)$ for $n =  L/8, 3L/16 (L=256) $ as defined in \eqref{eqn:corr_site_magplus} for $\alpha=1.1$}
\label{fig:MM_corrxt_3_unconserved}
\end{figure*}

\begin{figure*}[htp]
\centering
\begin{subfigure}{0.45\textwidth}
  \centering
  \includegraphics[width=\linewidth]{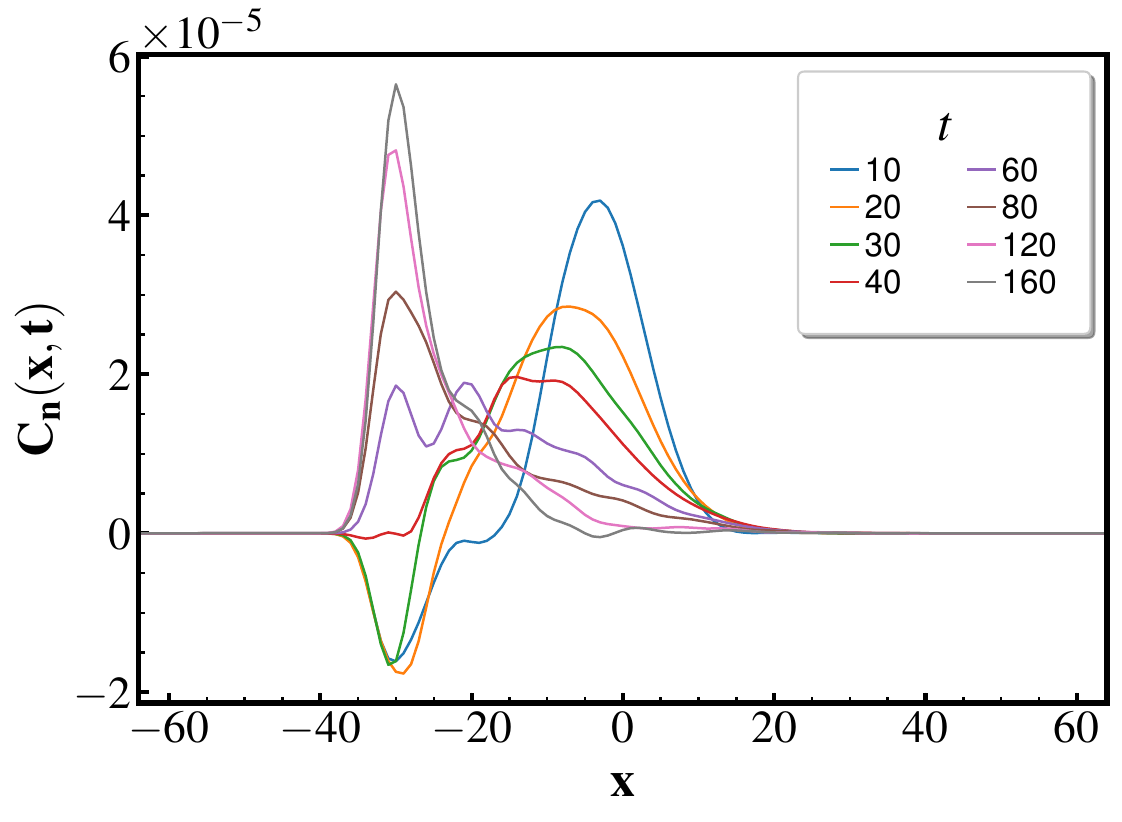}
  \caption{}
  \label{fig:CE_n_L4}
\end{subfigure}\hfill
\begin{subfigure}{0.45\textwidth}
  \centering
  \includegraphics[width=\linewidth]{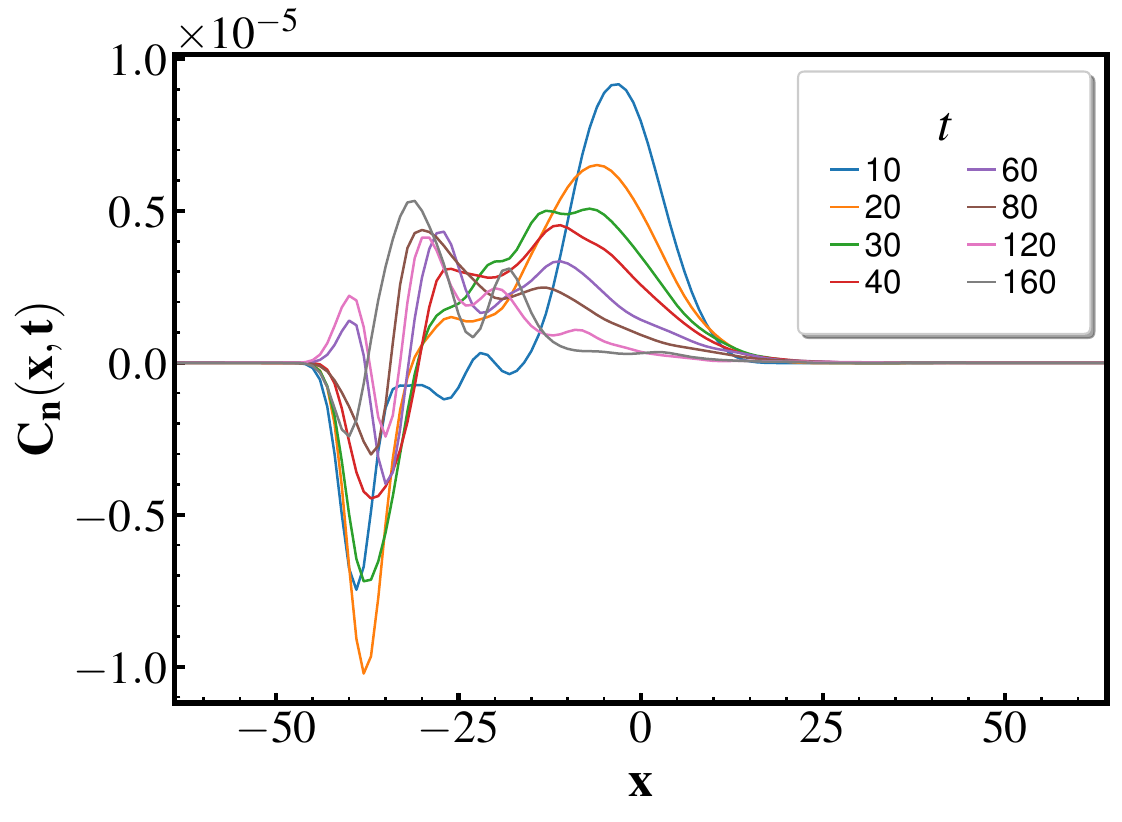}
  \caption{}
  \label{fig:CE_n_L2}
\end{subfigure}
\caption{\small Site-anchored correlator $C^{\mathcal{E}}_n(x,t)$ for $n = L/4, L/2$ ($L=256$) as defined in \eqref{eqn:corr_site_Ehyb} for $\alpha=1.1$.}
\label{fig:MM_corrxt_3_unconserved}
\end{figure*}

We calculate the site-anchored correlator of the energy-like scalar $\mathcal{E}_{n}$ as well, 
\begin{align}
\label{eqn:corr_site_Ehyb}
C^{\mathcal{E}}_{n}(x,t) &= \left \langle \frac{1}{\alpha^{n+x+1}}(\spin_{n+x} (t) \cdot \spin_{n+x+1} (t))\frac{1}{\alpha^{n+1}}(\spin_{n} (0) \cdot \spin_{n+1} (0))\right\rangle
\end{align}
Just like the previous case for $C_{\mathcal{M}}(x,t)$, it depends on the site index $n$, and $x$ runs from $-n+1$ to $L-n-1$.  Again, $C^{\mathcal{E}}_n(x,t)$ shows directionality in spreading. 
Despite these asymmetries at the individual level, the site-averaged correlator $C_{\mathcal{E}}(x,t)$ remains symmetric about $x =0$, similar to $C_{\magplus}(x,t)$, see Appendix \ref{App:symm_correlator}. An asymmetric site-anchored correlator is also obtained for the systems with $|\alpha|=1$ and open boundaries. However, these systems also admit periodic boundary conditions for which the site anchored correlator is always symmetric. This is in contrast to the the systems with $|\alpha| \neq 1$ which only admit open boundary conditions.

\section{Decorrelator and Chaos propagation} \label{sec:decorr}  
\begin{figure*}[htp]
\centering
\begin{subfigure}{0.49\textwidth}
  \centering
  \includegraphics[width=\linewidth]{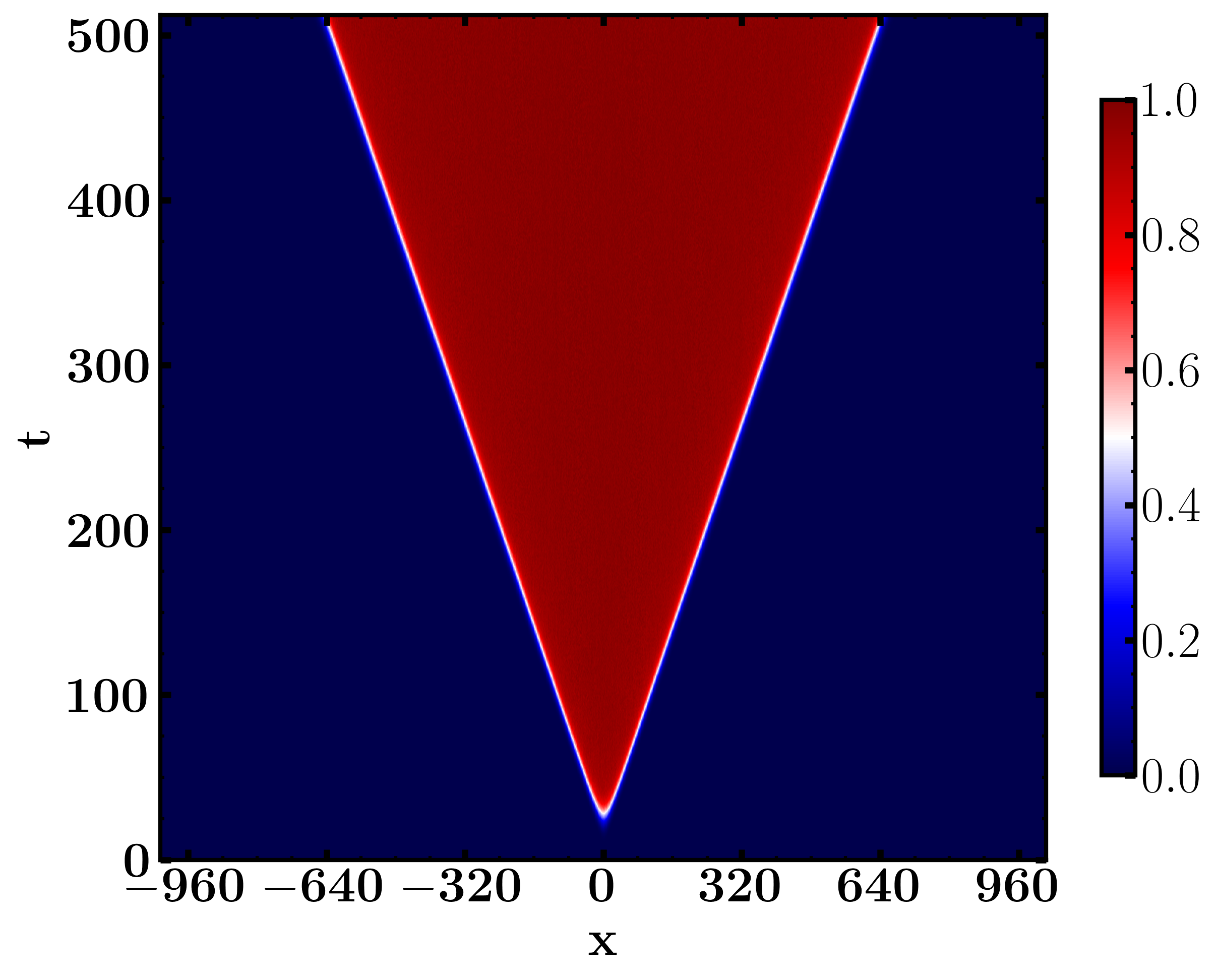}
  \caption{}
  \label{fig:puredrvn}
\end{subfigure}\hfill
\begin{subfigure}{0.49\textwidth}
  \centering
  \includegraphics[width=\linewidth]{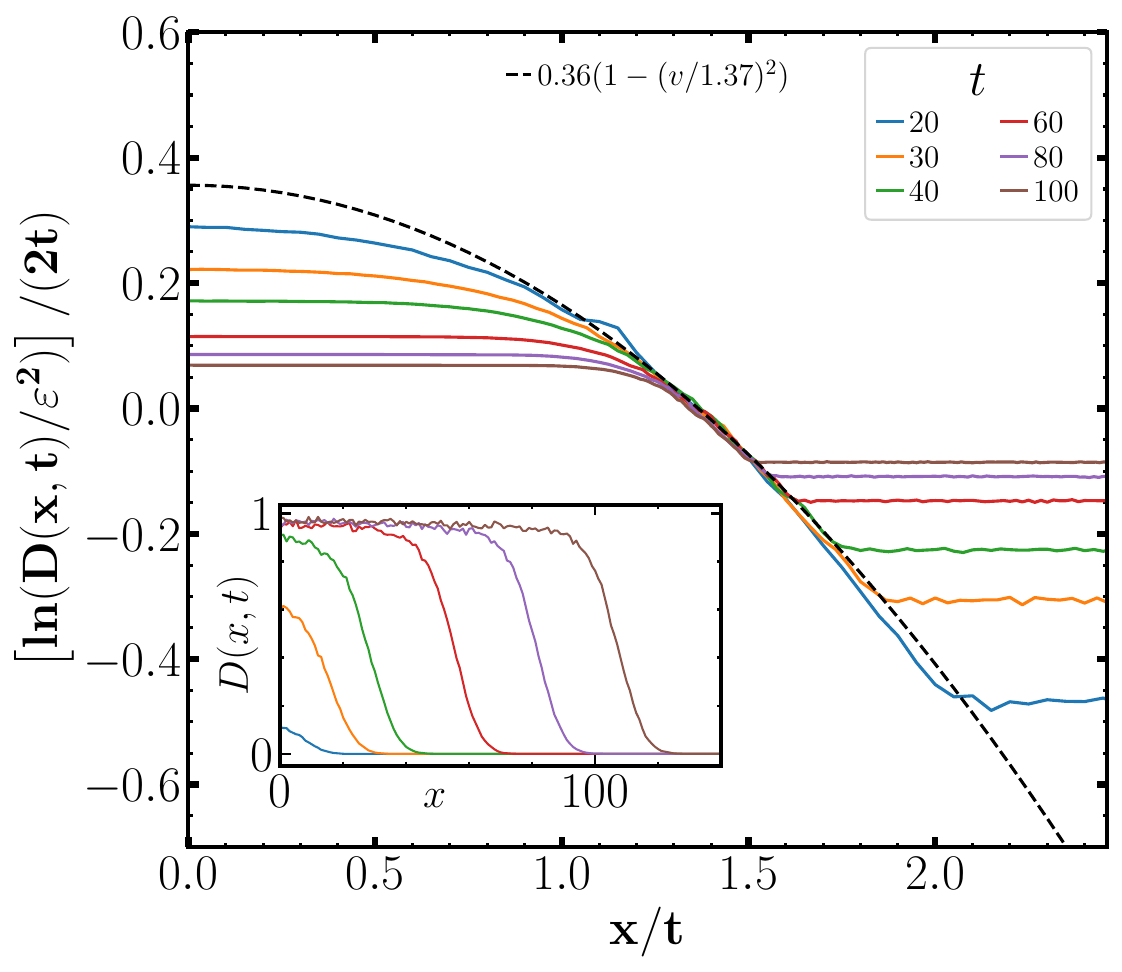}
  \caption{}
  \label{fig:logDxt_drvn}
\end{subfigure}
\caption{\small (Left) Colourmap of the decorrelator $D(x,t)$ calculated by averaging over pairs of initial conditions differing only at ${\bf S_0}$. This initial disturbance spreads ballistically, from which the butterfly velocity $v_B=1.37$ is obtained. (Right) The decorrelator given by the expression in \eqref{eqn:front} plotted as a function of $x/t$ for different values of $t$. There is a collapse of the curves in the vicinity of the front from which the Lyapunov exponent $\approx 0.36$ can be extracted. The inset shows $D(x,t)$ as a function of $x$ for different values of $t$. The existence of a front can be seen from the rapid decrease in the value of $D(x,t)$ as a function of $t$ (inset). 
  }
\label{fig:Dxt_logDxt_drvn}
\end{figure*}

We now turn to a different aspect of the dynamics of our system, namely the spreading of chaos. Chaotic dynamics is 
an important indicator of the presence of thermalization in Hamiltonian systems. Chaos spreading in the classical Heisenberg chain has been quantified through a numerical calculation of the decorrelator, which shows a ballistic spreading of chaos with a ``butterfly velocity''~\cite{das2018light}. We perform a similar calculation of the decorrelator for our system, focusing on the strictly antisymmetric case. 
The decorrelator is defined as
\begin{align}
    D( x, t) = 1 - \langle \spin_n^{(A)}( t)\cdot \spin_n^{(B)}( t) \rangle,
    \label{eqn:Decorr}
\end{align} 
where $(A)$ and $(B)$ are two different initial conditions which differ only at a single site of the lattice $n=0$, as 
$$\spin^{(B)}_0(0) = \spin^{(A)}_0(0) + \delta \spin_0 $$
where
\begin{align}
\begin{split}
\delta \spin_0 &= \varepsilon[\hat{\mathbf{n}} \times \spin^{(A)}_0(0)] \\
\hat{\mathbf{n}} &= \left[\hat{\mathbf{z}} \times \spin^{(A)}_0(0)\right]/\left | \hat{\mathbf{z}} \times \spin^{(A)}_0(0)
\right|
\end{split}
\label{eqn:Epsilonchange}
\end{align}
$\hat{\mathbf{z}}$ is the unit vector along an arbitrarily defined z-axis and $\varepsilon$ is the strength of the perturbation (\textit{i.e.} a measure of how different the two initial conditions are). Note that $$ \delta S_x^{\alpha}(t) \approx \dfrac{\partial S_x^{\alpha}(t)}{\partial S_0^{\beta}} \,\delta S_0^{\beta} 
    = \varepsilon n^{\gamma} \epsilon_{\beta \gamma \nu} S_0^{\nu}\dfrac{\partial S_x^{\alpha}(t)}{\partial S_0^{\beta}} 
    = \varepsilon n^{\gamma} \{ S_x^{\alpha}(t) , S_0^{\gamma}(0)\}$$
\begin{equation}
    \label{eqn:Decorr_poisson}
    D(x, t) = \frac{1}{2}\langle \delta \mathbf{S}(x,t)^2 \rangle \approx \frac{\varepsilon^2}{2} \langle \hat{\mathbf{n}} \cdot \{\mathbf{S}_x(t), \mathbf{S}_0 \}^2 \rangle
\end{equation}
Similar to the calculation of the two point correlators, all possible initial conditions are equally weighted while performing the average $\langle \dots \rangle$ in \eqref{eqn:Decorr}. 

We used $L=2048$, $\Delta t = 0.001-0.005$ in the numerical evolution of \eqref{eqn:drvn_dynmk} with RK4 integrator, averaged over 10000 uniformly drawn initial configurations, to evaluate the decorrelator. The spin arrays were stored with a quadruple numerical precision.  A colourmap of the decorrelator obtained is shown in Fig. (\ref{fig:Dxt_logDxt_drvn}). It can be clearly seen that the disturbance of strength $\varepsilon$ introduced at site $n=0$ spreads ballistically and symmetrically with a clearly defined butterfly velocity $v_B$. The slowest process in the global thermalization of the system is thus diffusive, yet information spreading is not inhibited by such slow dynamics and is ballistic
instead.

Chaos propagation in a many-body system such as ours cannot be merely captured by a temporal divergence. It was shown that systems can appear
chaotic{, apparently} without exhibiting a single positive Lyapunov exponent \cite{Deissler1987}. A velocity-dependent (or comoving) Lyapunov
exponent $\kappa(v)$ was defined to quantify the growth rate of a localized disturbance in a reference frame moving with a constant velocity $v$, 
such that the
deterministic dynamics were deemed chaotic when the maximum value of $\kappa(v)$ was positive.
The velocity of the comoving frame is analogous to the Lieb-Robinson velocity \cite{Lieb1972} of information scrambling in a many-body system with finite range interactions [for example, action of a local operator $\hat{\mathcal{A}}(x,t)$, on another local operator at a distant site $\hat{\mathcal{B}}(0,0)$].

We obtain a front velocity $v_B = 1.37(\pm 0.02) $ which is different from the value obtained for the classical Heisenberg model~\cite{das2018light} {reflecting} 
the difference in the microscopic dynamics of the two models. In the vicinity of the ballistically propagating front, the form of the decorrelator can be approximated as~\cite{das2018light} as
\begin{align}
\log
(D(x,t)/\varepsilon^2) &= 2 \kappa t [1 - (x/v_B t)^2],
\label{eqn:front}
\end{align}
where $\kappa$ is the Lyapunov exponent of the system. A fit to this form in the vicinity of the front is also shown in Fig.(\ref{fig:Dxt_logDxt_drvn}) from which one can obtain a Lyapunov exponent as well equal to $0.36(\pm 0.02)$. 

\subsection{Nearest neighbour nonreciprocal model: symmetries of the equation of motion and the decorrelator} \label{subsec:furtherNR_decorr}
\begin{figure*}[htp]
\centering
\begin{subfigure}{0.49\textwidth}
  \centering
  \includegraphics[width=\linewidth]{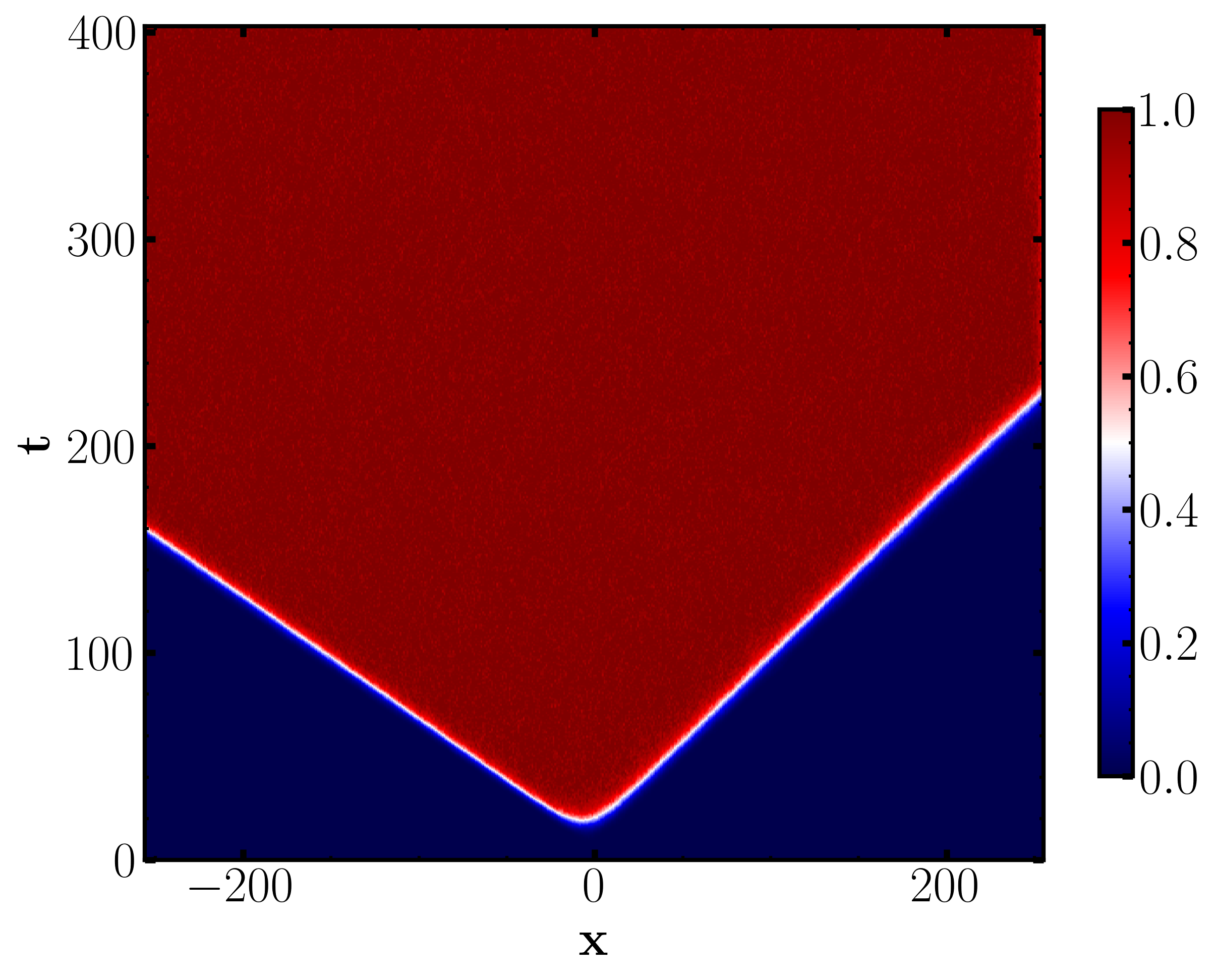}
  \caption{}
  \label{fig:genericnonreci}
\end{subfigure}\hfill
\begin{subfigure}{0.49\textwidth}
  \centering
  \includegraphics[width=\linewidth]{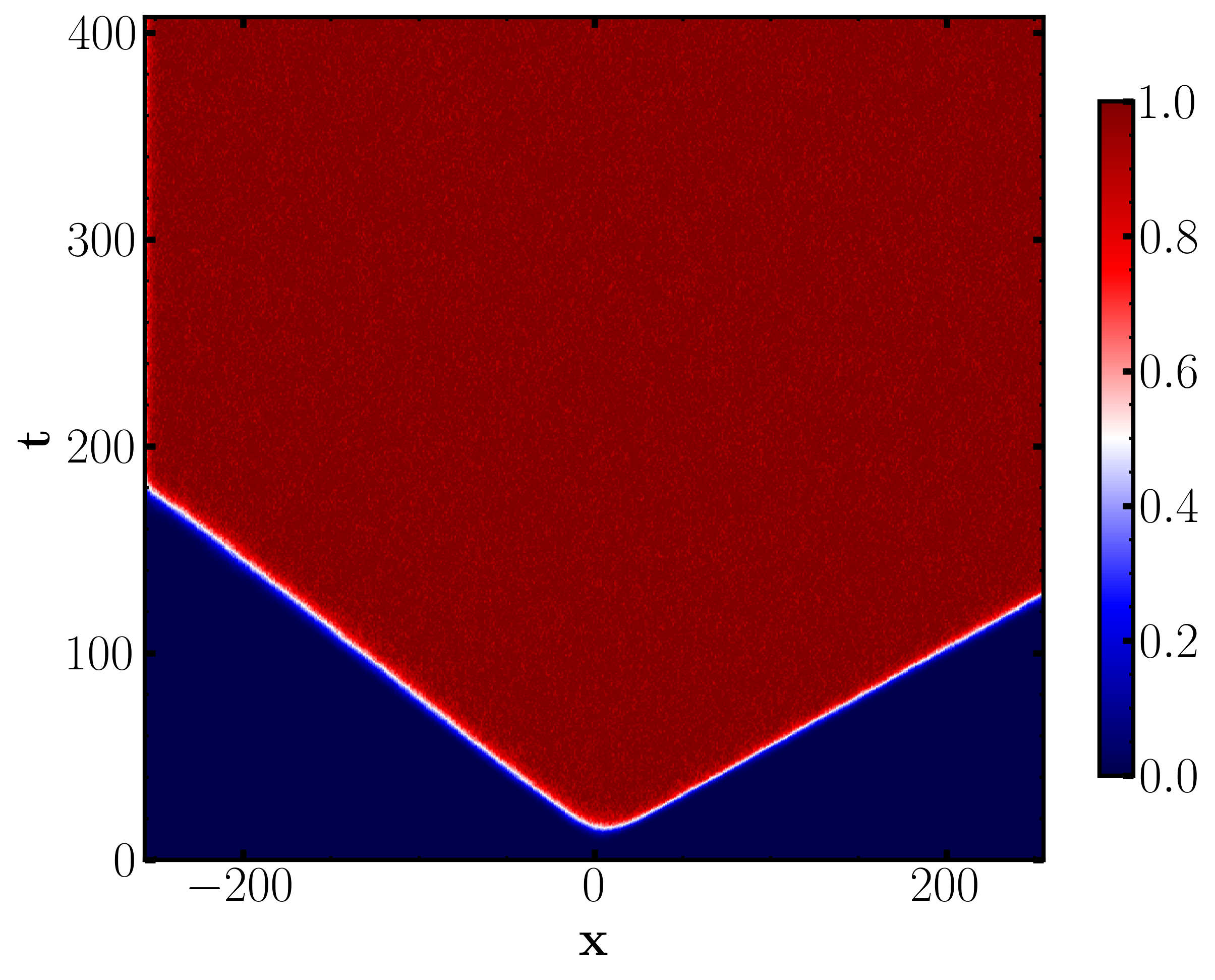}
  \caption{}
  \label{fig:Dxt_alpha125}
\end{subfigure}
\caption{\small  Colourmap of the decorrelator $D(x,t)$ calculated for 
$\alpha \neq 1$. The butterfly velocities for the left-right fronts are different: \textit{(Left)}  $\alpha = 0.80, v_{B_1}=1.86, v_{B_2} = 1.26$  \textit{(Right)} $\alpha = 1.25, v_{B_1} = 1.62, v_{B_2} = 2.33 $. 
  }
\label{fig:Dxt_3_hybrid}
\end{figure*}

\comnt{For clarity pertaining to the calculation below rewrite \eqref{eqn:hybrid_dynmk1} as
\begin{equation}
    \dot{\spin}_n = \spin_n \times (\lambda( \spin_{n+1} + \spin_{n-1}) + \mu( \spin_{n+1} - \spin_{n-1})) 
\end{equation}
denoting the explicit contribution of the Heisenberg and antisymmetric terms to the dynamics.}
Eqn. \eqref{eqn:hybrid_dynmk1} represents the dynamics governed by an interpolation of the symmetric and antisymmetric coupling strengths.
Our results on the decorrelator for these cases suggest that the rate of chaos propagation across the lattice, \textit{i.e.,} the butterfly speed, is affected by the nature of the coupling involved $(\alpha = \pm 1)$.
We see that when $|\alpha| \neq 1$, the reflection symmetry of the decorrelator about the site of initial perturbation ceases to exist, as shown in Fig.(\ref{fig:Dxt_3_hybrid}), and we observe two different slopes corresponding to two butterfly velocities. The butterfly velocities are  not symmetric with respect to the reflection $\alpha \rightarrow 1/\alpha$, which reverses the role of left-right neighbour interactions. On the other hand the time-ordered correlator remains unchanged when  $\alpha \rightarrow 1/\alpha$. 

\par 
 We now check for the invariance of the generalized equation of motion under symmetry transformations. Consider the generalized dynamics,
  \begin{align*}
  \tag{\ref{eqn:hybrid_dynmk1}}
  \dot{\spin}_x = \spin_x \times (\spin_{x+1} + \alpha \spin_{x-1}) .
  \end{align*}
  
Let $\invx$ be the spatial inversion operation on the lattice sites such that $\invx : \spin_x \mapsto \spin_{-x}$, or $\spin^{\invx}_x = \spin_{-x}$ for short. The equation of motion {\eqref{eqn:hybrid_dynmk1}} transforms under this operator as
\begin{align}
  \dot{\spin}^{\invx}_x = \spin^{\invx}_x	\times (\alpha \spin^{\invx}_{x+1} +  \spin^{\invx}_{x-1})
  \end{align}
{which is clearly not the same as \eqref{eqn:hybrid_dynmk1}}. 
To attempt to restore the invariance of the original equation, we {introduce} 
a further operation $\revertx$ 
\begin{align}
\label{eqn:operate_invert}
\dot{\spin}^{\revertx\invx}_x = \spin^{\revertx\invx}_x	\times (\alpha \spin^{\revertx \invx}_{x+1} + \spin^{\revertx \invx}_{x-1})
\end{align} 
{whose form is to be determined.}
Comparing with the original equation {\eqref{eqn:hybrid_dynmk1}}, $\revertx$ restores the invariance only when
\begin{align}
{\spin}^{\revertx\invx}_x =\alpha \spin_{-x} = \frac{1}{\alpha} \spin_{-x}
\end{align}
or $\alpha ^2 = 1$. The above condition gives us two familiar cases: the Heisenberg model, where $\spin^{\revertx\invx}_x = \spin_{-x}$, and the antisymmetric exchange model, where$\spin^{\revertx\invx}_x = -\spin_{-x}$, with the spins under the transformation operators leaving the equation of motion invariant.
\par When $\alpha = \pm 1$, \eqref{eqn:operate_invert} implies that $\spin^{\revertx\invx}_x (t) = \pm \spin_{-x} (t)$ as we integrate the equation from the initial conditions $\spin^{\revertx\invx}_x (t) = \spin^{\revertx \invx}_x(0)$. 
 This means that two distinct initial configurations, say  $(A)$ and $(C)$, are invariant under $\revertx\invx$ and 
\begin{align*}
\spin_x^{(A)}(t) = \spin_{-x}^{(C)}(t) \Leftarrow \spin_x^{(A)}(0) = \spin_{-x}^{(C)}(0), \, \rm{for} \,\, \alpha = 1
\end{align*}
and 
\begin{align*}
\spin_x^{(A)}(t) = -\spin_{-x}^{(C)}(t) \Leftarrow \spin_x^{(A)}(0) = -\spin_{-x}^{(C)}(0), \, \rm{for} \,\, \alpha = -1
\end{align*}
This one-one mapping translates to the definition of the decorrelator 
\begin{align}
\label{eqn:symm_decorr1}
\spin^{(A)}_x(t)\cdot \spin^{(B)}_x(t) = \spin^{(C)}_{-x}(t)\cdot \spin^{(D)}_{-x}(t)
\end{align}
since 
\begin{align*}
\spin^{(A)}_x(0) = \spin^{(C)}_{-x}(0) \implies \spin^{(B)}_x(0) = \spin^{(D)}_{-x}(0), \quad ( \alpha = \beta)
\end{align*} 
where (B) and (D) are the perturbed copies of the configurations (A) and (C) respectively. A similar argument for $\alpha = -1$ 
{leaves} \eqref{eqn:symm_decorr1} unchanged. 
For $\alpha = 1$ the initial conditions also satisfy
$\spin^{(C)}_x(0)= \spin^{(D)}_x (0)$ for all $ x \neq 0$
and 
\begin{align*}
\begin{split}
\spin^{(D)}_0(0) &= \spin^{(C)}_0(0) + \varepsilon(\hat{\mathbf{n}} \times \spin^{(C)}_0(0)) \\
\delta \spin^{(A)}_0 &= \delta \spin^{(C)}_0
\end{split}
\end{align*}
 averaging over all such pairs of initial conditions,
\begin{align}
\begin{split}
\langle \spin^{(A)}_x(t)\cdot \spin^{(B)}_x(t) \rangle &= \langle \spin^{(A)}_{-x}(t)\cdot \spin^{(B)}_{-x}(t)\rangle \\
\implies D_x(t) = D_{-x}(t)
\end{split}
\end{align} 
 For $\alpha = -1$ the difference between the pair of initial conditions goes as
\begin{align*}
\spin^{(D)}_0(0) &= \spin^{(C)}_0(0) - \varepsilon(\hat{\mathbf{n}} \times \spin^{C}_0(0))
\end{align*}
 However, since the decorrelator goes as $\varepsilon^2$ \eqref{eqn:Decorr_poisson}, the product $\spin^{(A)}_0(0) \cdot (\spin^{(A)}_0(0) \pm \delta \spin^{(A)}_0)$ is unaffected by the sign in front of $\varepsilon$.
Hence, we find $D_x(t) = D_{-x}(t)$ here as well. Since such a transformation does not exist for the generic nonreciprocal model, the decorrelator is asymmetric with respect to the site of initial perturbation.


\par We now look at the trend in the antisymmetric exchange model with next-nearest-neighbour coupling.
\begin{figure}[h!]
    \begin{minipage}[t]{0.64\textwidth} 
        \begin{minipage}[t]{0.5\linewidth} 
            \centering
            \includegraphics[width=0.95\linewidth]{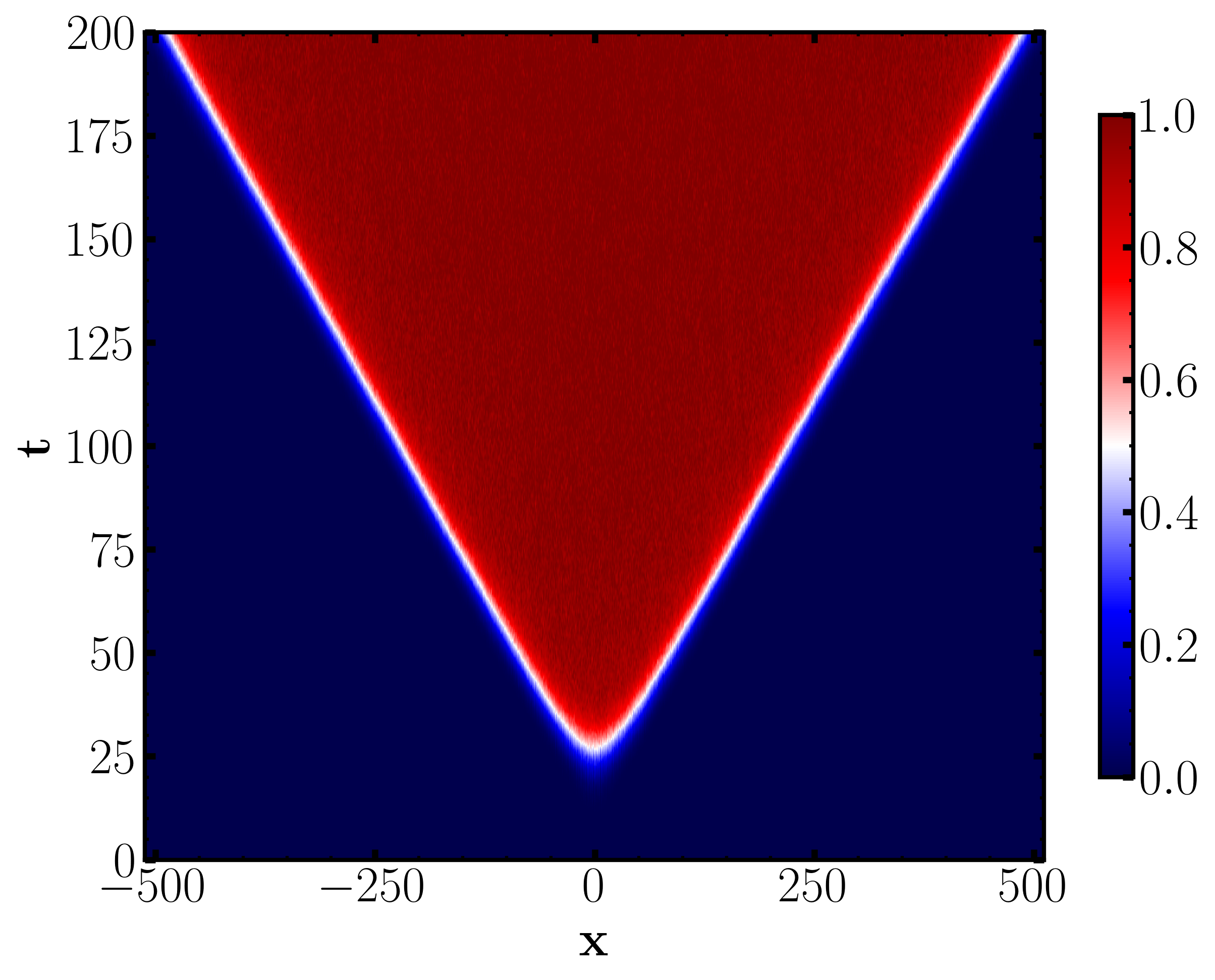}
            \label{fig:fig1_tl}
        \end{minipage}\hfill
        \begin{minipage}[t]{0.5\linewidth}
            \centering
            \includegraphics[width=0.95\linewidth]{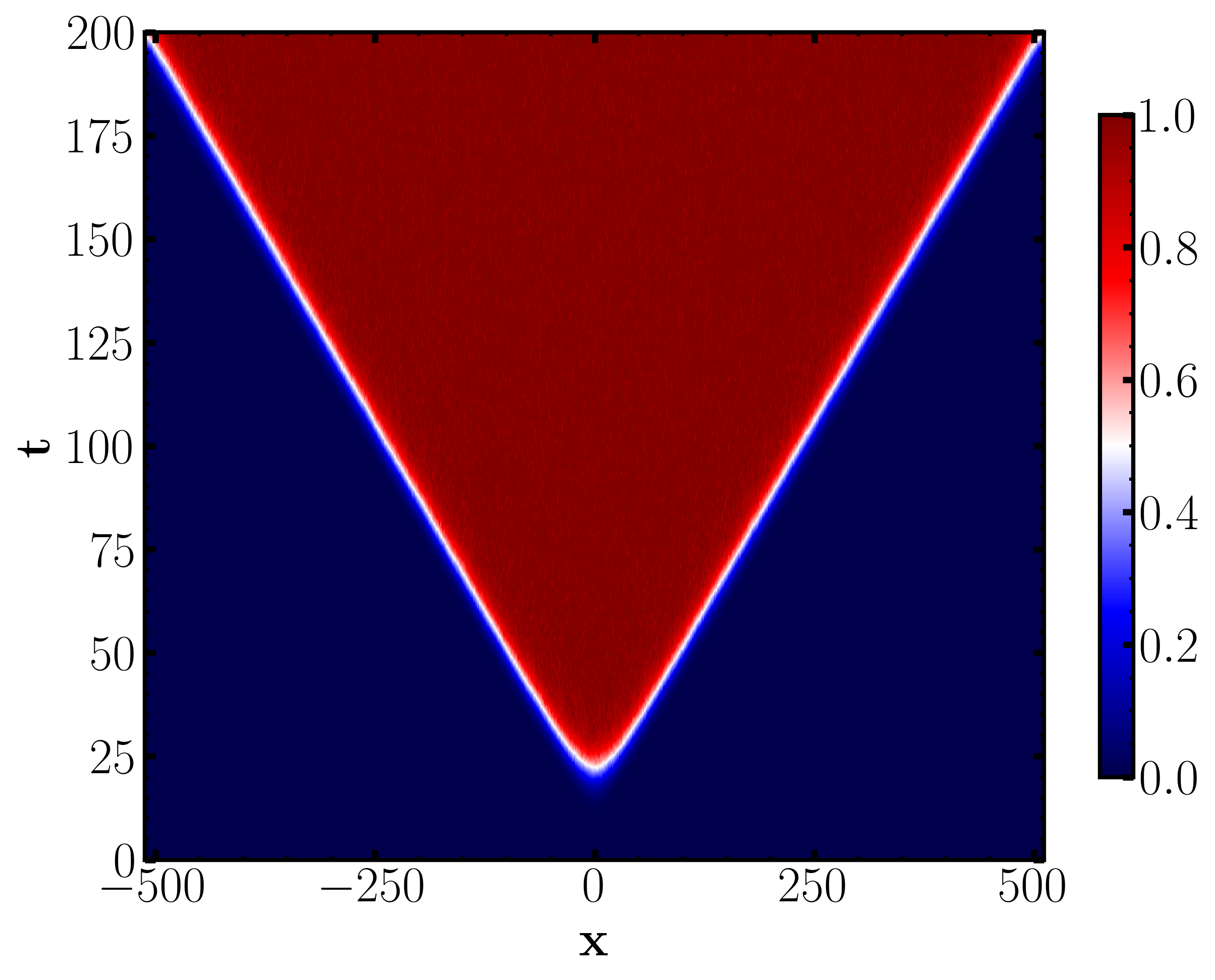}
            \label{fig:fig2_tr}
        \end{minipage}
        \vspace{1em} 

        \begin{minipage}[t]{0.5\linewidth}
            \centering
            \includegraphics[width=0.95\linewidth]{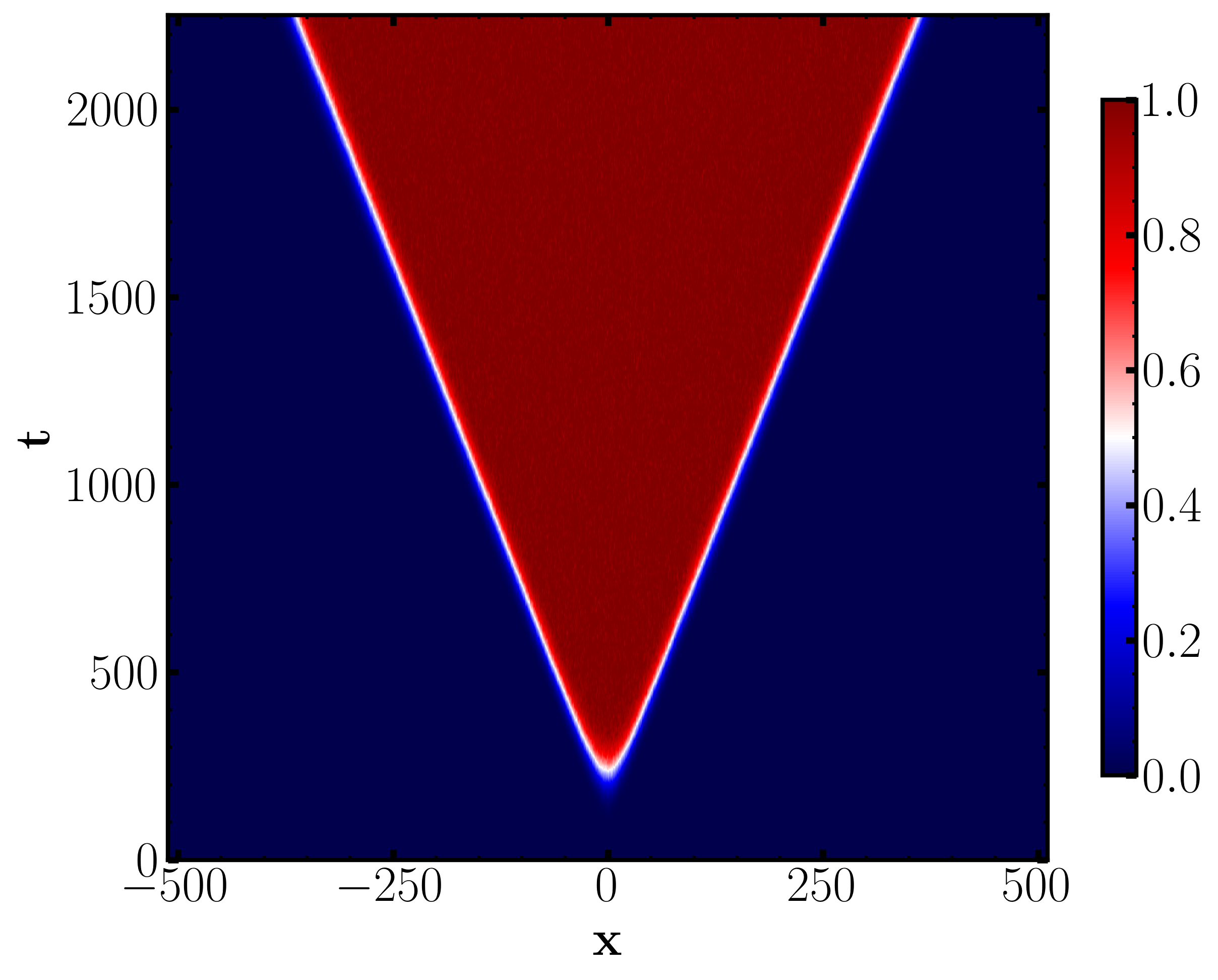}
            \label{fig:fig3_bl}
        \end{minipage}\hfill
        \begin{minipage}[t]{0.5\linewidth}
            \centering
            \includegraphics[width=0.95\linewidth]{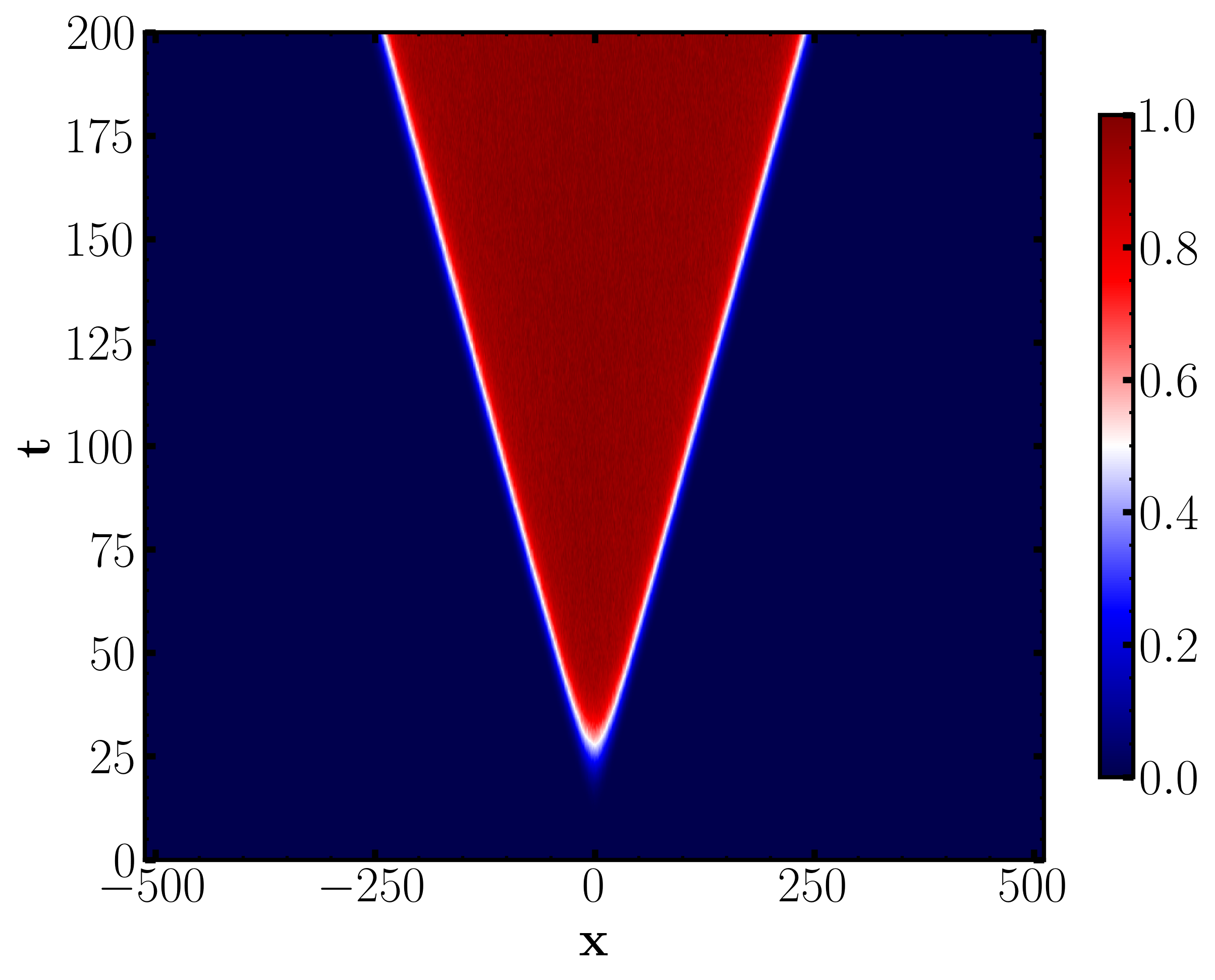}
            \label{fig:fig4_br}
        \end{minipage}
    \end{minipage}\hfill 
    \begin{minipage}[c]{0.35\textwidth} 
        \centering
        \vspace{5em}
        \includegraphics[width=6cm, height=6cm]{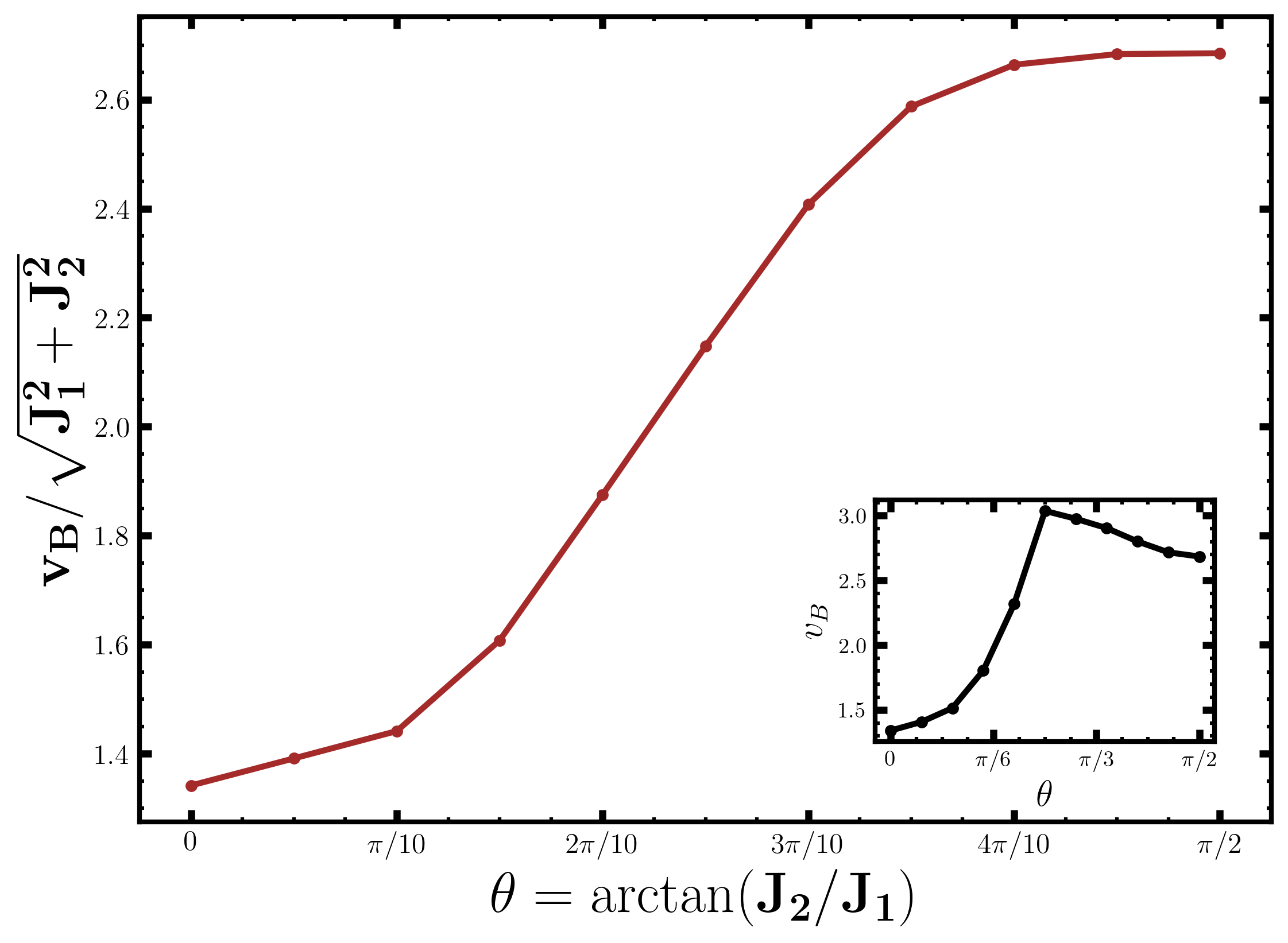}
        \label{fig:fig5_rs}
    \end{minipage}
    \caption{\small{\textit{(Left)} Decorrelator plots with decreasing next-nearest-neighbour interaction strength, with $J_1/J_2 = 0.1, 0.25$ in the \textit{(Top)}, subfigures and $J_2/J_1 = 0.04, 0.016$ in the \textit{(Bottom)} subfigures respectively. \textit{(Right)} Butterfly velocity trend: $v_B$ shows a monotonic increase with an increasing $J_2/J_1$ interaction , depicted as $\tan^{-1}(J_2/J_1)$, after being scaled by a factor of $\sqrt{J_1^2 + J_2^2}$. }}
    \label{fig:Dxt_NNN} 
\end{figure}

\subsection{Next-nearest-neighbour interaction model} {We showed in section \ref{subsec:further} that the model ceases to be Hamiltonian upon extending the antisymmetric exchange to the next-nearest neighbour spins,} as there are no conserved (scalar) quantities upto the second  order in $\{\spin_n\}$. {The equation of motion for the model is}
\begin{equation}
    \dot{\spin}_n = J_1 \spin_n \times (\spin_{n+1} - \spin_{n-1} ) + J_2 \spin_n \times (\spin_{n+2} - \spin_{n-2})
    \label{eqn:nnnbr_dnmk}
\end{equation}
with periodic boundary conditions on $n \in \{1,2,..., L\}, \, L = 2m, \, m \in \naturalset$.
{In the simple limiting cases 
$J_1 , J_2 = (1, 0)$ and $(0,1)$, \eqref{eqn:nnnbr_dnmk} reduces, respectively, to 
Eq. \eqref{eqn:drvn_dynmk}, and 
to a purely antisymmetric model on the even 
sites of a bipartite lattice system. 
The butterfly velocity of the decorrelator front for the latter case is trivially twice that of 
the original nearest-neighbour model.} 

While studying the correlator for such a model without conserved quantities is {not very informative,} 
the study of the decorrelator front reveals an interesting pattern (\ref{fig:Dxt_NNN}). In the 
case of 
{negligible} nearest-neighbour interaction and strong next-nearest-neighbour interaction, the decorrelator effectively propagates on only one of the sublattices, where the initial perturbation was added. As the relative interaction strength $J_1/J_2$ is increased, the sublattices communicate 
{the growth of the initial perturbation to each other,} and the decorrelator eventually propagates across all the sites. 
The propagation front velocity appears to be the greatest for $J_1=J_2$ (inset of the \textit{(Right)} subfigure in Fig.\ref{fig:Dxt_NNN}). 
The velocity depends on the magnitudes of $J_1$ and $J_2$. When it is scaled by the factor $(J_1^2 + J_2^2)^{1/2}$, it changes monotonically with $J_1/J_2$. This is consistent with the expectation that the fastest propagation is for the case with $J_1 = 0 \text{ but } J_2 \neq 0$, since the disturbance jumps two lattice spacings in each step in this case.
We find that the velocity of the chaos propagation front decreases monotonically from $v_B = 2.69$ at $J_1/J_2 \rightarrow 0$ to $v_B = 1.35$ at $J_2/J_1 \rightarrow 0$, scaled by the factor $(J_1^2 + J_2^2)^{1/2}$ where all values of $J_1, J_2$ lie within the range $[0,1]$.

\section{Conclusion}\label{sec:Conclusion} Our study of the classical Heisenberg spin chain with nonreciprocal interactions uncovers a hidden symplectic structure that renders the generic nearest-neighbour model effectively reciprocal when expressed in terms of transformed spin variables. In this representation, the total magnetization and energy are conserved, 
{as confirmed} by our numerical studies. The antisymmetric coupling model emerges as a special case where these conservation laws persist even under periodic boundary conditions, the passage to infinite system size poses no difficulty, and the densities of the conserved quantities diffuse. {Except in that special case, the transformation to reciprocal form comes at a price: the local non-centrosymmetry of the original formulation is amplified to an exponential dependence of local properties on position, and a nonexistent thermodynamic limit.} Extending the interaction range beyond nearest neighbours breaks these conservation laws, with certain exceptions. 
Chaos propagation remains ballistic across the nonreciprocal models, with a symmetric decorrelator cone observed only in the antisymmetric coupling case.
Investigations of the nature of the long-time statistical steady state and issues regarding the thermodynamic limit in the 
the nearest-neighbour nonreciprocal model are fertile directions for future work.

\paragraph{Acknowledgements:}
NB was supported by a scholarship from the UGC, India. PD acknowledges support from Kishore Vaigyanik Protsahan Yojana (KVPY) from DST, Govt. of India. SM acknowledges a grant from the DST, Govt. of India. SR thanks the ANRF, India, for support in the form of a J C Bose National Fellowship (till Nov 2025) and a National Science Chair thereafter. The authors would like to thank Subhro Bhattacharjee for productive discussions, and for pointing out the conserved nature of staggered magnetization in our model. We also thank an anonymous referee of an earlier version of this paper who pointed out that the antisymmetric model {can be transformed to Hamiltonian form.} 


\appendix
\section{{Liouville Theorem} for nearest-neighbour nonreciprocal spin dynamics}
\renewcommand{\theequation}{A.\arabic{equation}}
\setcounter{equation}{0}
Consider the dynamics governed by the nonreciprocal interaction $(J_{n, n+1} = -J_{n+1,n}) $ :
\begin{align}
\label{eqn:Liouville1_negv1}
    \dot{S}^{\mu}_n &= \epsilon_{\mu \nu \sigma}(S^{\nu}_n S^{\sigma}_{n+1} + \alpha S^{\nu}_n S^{\sigma}_{n-1})
\end{align}

Its divergence in the spin space is :
\begin{align}
\label{eqn:Liouville2_divergfree}
\begin{split}
     \sum_{n, \mu} \dfrac{\partial \dot{S}_n^{\mu}}{\partial S_n^{\mu}} &= \sum_{n, \mu} \dfrac{\partial}{\partial S^{\mu}_n} \left[\sum_{\nu \sigma} \epsilon_{\mu \nu \sigma}(S^{\nu}_n S^{\sigma}_{n+1} + \alpha S^{\nu }_n S^{\sigma}_{n-1}) \right] \\ 
     &= \sum_{n, \alpha} \left[ \sum_{\nu \sigma}\epsilon_{\mu \nu \sigma} \delta_{\mu \nu} (S^{\sigma}_{n+1} + \alpha S^{\sigma}_{n-1}) \right] \\
     &= 0
\end{split}
\end{align}

The total time derivative of the density of spin configurations is
\begin{align*}
\label{eqn:Liouville3_totalDrho}
\begin{split}
  \dfrac{d \rho (\{ \spin_n \}, t) }{d t}  
  &= \dfrac{\partial \rho}{\partial t} + \sum_{n, \mu} \dfrac{\partial \rho}{\partial S^{\mu}_n} \dot{S}^{\mu}_n 
  \end{split}
\end{align*}

Continuity equation of the phase space density gives us,
\begin{equation*}
    \dfrac{\partial \rho}{\partial t} = -\sum_{n, \mu}\dfrac{\partial (\rho \dot{S}_n^{\mu})}{\partial S_n^{\mu}}
    =  - \sum_{n, \mu} \left[ \dfrac{\partial \rho }{\partial S^{\mu}_n} \dot{S}^{\mu}_n + \rho \dfrac{\partial \dot{S}_n^{\mu}}{\partial S_n^{\mu}}\right]
    =  - \sum_{n, \mu}  \dfrac{\partial \rho }{\partial S^{\mu}_n} \dot{S}^{\mu}_n \quad \quad \quad \text{(using \ref{eqn:Liouville2_divergfree})}
    \label{eqn:Liouville4_partialdrho}
\end{equation*}

Thus we see that 
\begin{align}
\label{eqn:Liouvillethm_main}
    \dfrac{d \rho (\{ \spin_n \}, t) }{d t} &= 0, 
\end{align}
{which can be termed a Liouville theorem} for the nearest neighbour, generic nonreciprocal Heisenberg model.

A similar argument was presented in the work by Hanai \cite{Hanai2024} on nonreciprocal spin models, starting with the Landau-Lifshitz equation.

\section{{Continuity equations}}
\label{App:continuity}
\renewcommand{\theequation}{B.\arabic{equation}}
\setcounter{equation}{0}
{The results for the continuity equations \eqref{eqn:contnutMN_continuum}- \eqref{eqn:antisymm_Edyn} can be proven with a little bit of algebra and coarse-graining. 
Defining the local magnetization $\magg_n$ and staggered magnetization $\stagg_n$ at the bond connecting the spins $\spin_{2n}$ and $\spin_{2n+1}$,}
\begin{equation}
\label{eqn:defMN}
\magg_n = (\spin_{2n} + \spin_{2n+1})/2 , \quad \quad 
\stagg_n  = (\spin_{2n} - \spin_{2n+1})/2,
\end{equation}
{and  the forward difference operator as $\dd \mathbf{X}_r = \mathbf{X}_{r+1} - \mathbf{X}_r$, the equations of motion for the odd and even site spins}
 \begin{align}
 \label{eqn:motion}
 \begin{split}    
 \dot{\spin}_{2n} &= \spin_{2n} \times (\spin_{2n+1} - \spin_{2n-1}) \\
 &= (\magg_n + \stagg_n) \times (\dd \magg_{n-1} - \dd \stagg_{n-1}) \\
\dot{\spin}_{2n+1} &= \spin_{2n+1} \times (\spin_{2n+2} - \spin_{2n}) \\
&= (\magg_n - \stagg_n) \times (\dd \magg_n + \dd \stagg_n).
\end{split}
\end{align}
{Next, we write down the equations of motion for $\magg_n$ and $\stagg_n$:}
\begin{align}
\label{eqn:Mdyn_appdx}
\begin{split}
        \dot{\magg}_n = \half(\dot{\spin}_{2n} + \dot{\spin}_{2n+1}) &= \frac{1}{2} \Bigl[ \magg_n \times (\dd \magg_{n-1} + \dd \magg_n) - \magg_n \times (\dd \stagg_{n-1} - \dd \stagg_n) \\
    &+ \stagg_n \times (\dd \magg_{n-1} - \dd \magg_n) - \stagg_n \times (\dd \stagg_{n-1} + \dd \stagg_n) \Bigr] 
\end{split}
\end{align}
\begin{align}
\label{eqn:Ndyn_appdx}
\begin{split}
        \dot{\stagg}_n = \half(\dot{\spin}_{2n} - \dot{\spin}_{2n+1}) &=  \frac{1}{2} \Bigl[ \magg_n \times (\dd \magg_{n-1} - \dd \magg_n) - \magg_n \times (\dd \stagg_{n-1} + \dd \stagg_n) \\
    & + \stagg_n \times (\dd \magg_{n-1} + \dd \magg_n) - \stagg_n \times (\dd \stagg_{n-1} - \dd \stagg_n) \Bigr]
\end{split}
\end{align}

\paragraph{Pseudo-Energy Current ($\mathcal{J}^{\Epseudo}$) and continuity equation: }Defining the pseudo-energy symmetric to the even site,
    $\Epseudo_n = -\frac{1}{2} \spin_{2n} \cdot (\spin_{2n+1} - \spin_{2n-1})$,
and noting that the local field dynamics leaves the central spin evolution unchanged ($\dot{\spin}_k \cdot {\spin}_k = 0$), the time derivative simplifies to $\dot{\Epseudo}_n = -\frac{1}{2} \spin_{2n} \cdot (\dot{\spin}_{2n+1} - \dot{\spin}_{2n-1})$. Substituting the odd-site equations of motion:
\begin{equation}
    \dot{\Epseudo}_n = -\frac{1}{2} (\magg_n + \stagg_n) \cdot \left[ (\magg_n - \stagg_n) \times (\dd \magg_n + \dd \stagg_n) - (\magg_{n-1} - \stagg_{n-1}) \times (\dd \magg_{n-1} + \dd \stagg_{n-1}) \right]
\end{equation} 
{The discrete evolution of $\Epseudo_n$ naturally forms a flux difference, allowing us to identify the exact discrete energy current:}
\begin{equation}
    \mathcal{J}^{\Epseudo}_n = \frac{1}{2} (\magg_n + \stagg_n) \cdot \left[ (\magg_n - \stagg_n) \times (\dd \magg_n + \dd \stagg_n) \right]
\end{equation}
{In the continuum limit, the pseudo-energy becomes
$\Epseudo = -\frac{a}{2} (\magg + \stagg) \cdot \partial_x (\magg - \stagg)$. 
Replacing the differences with $ \partial_x$, and  absorbing the lattice spacing $a$ into the time, the pseudo-energy current becomes:}
\begin{equation}
\label{eqn:curr_epseudo}
    \begin{split}
    \mathcal{J}^{\Epseudo} &=  \half (\magg + \stagg) \cdot \left[ (\magg - \stagg) \times \partial_x (\magg + \stagg) \right] \\
    &= \half \partial_x(\magg+ \stagg) \cdot \left[ (\magg + \stagg) \times (\magg - \stagg) \right]\\
    &=  -(\magg \times \stagg) \cdot \partial_x (\magg + \stagg)
    \end{split}
\end{equation}
{
Thus,
\begin{align}
\tag{\ref{eqn:antisymm_Edyn}}
    \partial_t \Epseudo 
    + \partial_x [ (\stagg \times \magg) \cdot \partial_x (\magg + \stagg)] = 0
\end{align}
}
\subsection{Hydrodynamic modes and transient fast modes} \label{subsec:modes}
\begin{figure}[H]
\centering
\includegraphics[scale=0.36]{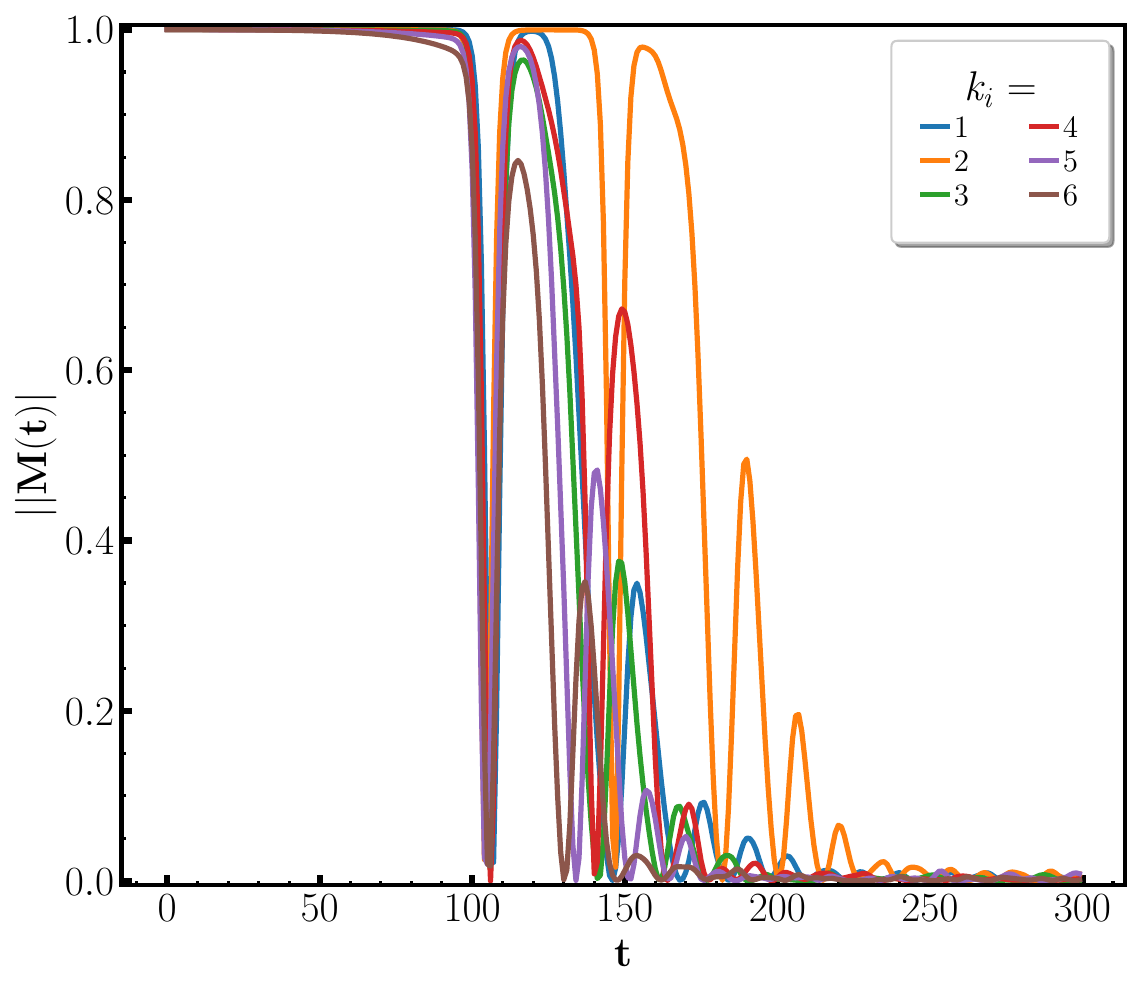}
\caption{\small The evolution of $|\magg(t)| = |\sum_n {\bf S}_n(t)|$ with time for different initial states of the form ${\bf S}_n(t=0) = \delta M \cos(2 \pi k n/L) \hat{e}_1 + \delta M \sin(2 \pi k n/L) \hat{e}_2 + M \hat{e}_3$. These represent states with a uniform value $M$ of one component of the spin with a modulation of strength $\delta M$ and wavevector $k$ of the other two components. The initial states thus have a total non-zero magnetization. It can be seen that this magnetization decays to zero as a function of time consistent with a non-zero decay time. The features (bounces) in the early time behavior are transients that cannot be captured by the hydrodynamic theory.}
\label{fig:Mkt_decayvs_t}
\end{figure}
Equations \eqref{eqn:contnutMN_continuum}-\eqref{eqn:antisymm_Edyn} have been obtained for the hydrodynamic (\textit{i.e.} low wavenumber) modes of $\magg$ and $\stagg$ in purely antisymmetric exchange model. The conservation law for $\stagg$ suggests that it is a slow variable evolving against a backdrop of the decaying non-conserved variable $\magg$. We expect that the effect of the large wavenumber (fast) modes on these hydrodynamic modes is to provide a noise for each, which in conjunction with the non-linearities in Eqn. \eqref{eqn:contnutMN_continuum}
give rise to diffusion for the conserved mode $\stagg$ and relaxation for the non-conserved mode $\magg$. 
We can check that the diffusive behaviour in $\stagg$-dynamics can be obtained, to a first order approximation, from the steady state description of $\magg $: 
\begin{align*}
\partial_t{\magg}(x,t) = 0 \quad \implies \magg \approx -\tau (\stagg \times \partial_x \stagg),
\end{align*}

\begin{align}
    \label{eqn:Diffusion_stagg}
    \begin{split}
     \partial_t{\stagg} &= \partial_x(\stagg \times \magg) \\ 
              &\approx -\tau \partial_x (\stagg \times (\stagg \times \partial_x \stagg)) \\ 
    &= \tau \partial_x (\stagg^2(\mathbf{1} - \hat{\stagg}\hat{\stagg} \cdot ) \partial_x \stagg) 
    \end{split}
\end{align}
{ The behaviour of the fast variable ${\bf M} = \sum_n {\bf S}_n$ can be studied starting from an initial state with ${\bf M} \neq 0$. Note that perfectly aligned (e.g. ${\bf S}_n = \hat{\bf e}_3 \, \forall \,n$) or perfectly alternating (say ${\bf S}_n = (-1)^n\hat{\bf e}_3 \, \forall \, n$) configurations do not evolve at all, $\dot{\bf S}_n \equiv 0$, and therefore trivially $(d/dt)\sum_n {\bf S}_n=0$ for such initial conditions. We therefore choose a generic nearly uniform state, in the form of a small periodic modulation added to a perfectly aligned state, i.e. ${\bf S}_n(t=0) = \delta M \cos(2 \pi k n/L) \hat{e}_1 + \delta M \sin(2 \pi k n/L) \hat{e}_2 + M \hat{e}_3$. It can be seen from Fig.(\ref{fig:Mkt_decayvs_t}) that the total magnetization ${\bf M}$ decays fairly rapidly to zero for different values of the modulation wavenumber $k$. Further, the relaxation of the magnetization in the limit $k \rightarrow 0$ in a finite time is consistent with our assumption of a finite $\tau$ arising from the hydrodynamics. This can be seen in the form of a fairly rapid relaxation of the magnetization for the lowest value of $k(=1)$ possible in our numerics. The complicated early time behavior of the magnetization, which shows bounces as a result of the precessional dynamics, is beyond the scope of the hydrodynamic framework.}
\comnt{
A numerical result showing the rapid decay of the magnetization in the antisymmetric exchange model for different initial states, is shown in Fig.(\ref{fig:Mkt_decayvs_t}). 
In particular, for initial conditions in which the magnetization has a modulation at a wavevector $k$ about a uniformly magnetized state, the relaxation time is finite in the limit $k \rightarrow 0$ as expected due to the finite value of $\tau$ we have assumed in our theory. \\
The behaviour of the non-conserved magnetization is captured in the transient state. We let the purely nonreciprocal spin dynamics evolve for initial states with a non-zero magnetization: a small periodic modulation of the spins is added to a perfectly aligned state. It can be seen from Fig.(\ref{fig:Mkt_decayvs_t}) that it decays fairly rapidly to zero for different values of the modulation wavenumber $k$. Further, the relaxation of the magnetization in the limit $k \rightarrow 0$ in a finite time is consistent with our assumption of a finite $\tau$ arising from the hydrodynamics. This can be seen in the form of a fairly rapid relaxation of the magnetization for the lowest value of $k(=1)$ possible in our numerics. A description of the structure (bounces) in the early time behavior of the magnetization is beyond the scope of the hydrodynamic framework.
}

\section{Spatially symmetric correlators of the linear and scalar bilinear quantities in $\magplus$}\label{App:symm_correlator}
\renewcommand{\theequation}{C.\arabic{equation}}
\setcounter{equation}{0}

The spatial spread of the total site-averaged correlation function $C_{\magplus}(x,t)$ is symmetric about $x=0$. First we will prove that $C_{\magplus}(x,t) = C_{\magplus}(x,-t)$. If we invert the time i.e. $t \to -t$ and also make the substitution ${\tilde{\spin}_n(t)} = -\spin_n(-t)$, then,
    \begin{equation}
    \begin{split}
        \frac{d\spin_n(-t)}{-dt} &= \spin_n(-t) \times \left(\spin_{n+1}(-t) + \alpha \spin_{n-1}(-t)\right)\\
        \implies \frac{d\bm{\tilde{S}}_n(t)}{dt} &= \bm{\tilde{S}}_n(t) \times \left(\bm{\tilde{S}}_{n+1}(t) + \alpha \bm{\tilde{S}}_{n-1}(t)\right)\\
    \end{split}
\end{equation}
These substitutions leave the form of \eqref{eqn:hybrid_dynmk1} invariant. Hence,
        \begin{equation}
    \begin{split}
        C_{\magplus}(x,t) &=\left\langle \mathop{\sum\nolimits'}_n\frac{1}{\alpha^{n+x}}\bm{\tilde{S}}_{n+x} (t) \cdot \frac{1}{\alpha^{n}}{\tilde{\spin}}_n (0)\right\rangle\\
        &= \mathop{\sum\nolimits'}_n\left\langle\frac{1}{\alpha^{n+x}}(-\spin_{n+x} (-t)) \cdot \frac{1}{\alpha^{n}}(-\spin_n (0))\right\rangle\\
         &= \mathop{\sum\nolimits'}_n\left\langle\frac{1}{\alpha^{n+x}}\spin_{n+x} (-t) \cdot \frac{1}{\alpha^{n}}\spin_n (0)\right\rangle\\
    \implies C_{\magplus}(x,t) &= C_{\magplus}(x,-t)
    \end{split}
    \end{equation}
    Now we will prove that $C_{\magplus}(x,t) = C_{\magplus}(-x,-t)$.
    Without loss of generality, consider $x> 0$. Then,
    \begin{equation}
    \begin{split}
    C_{\magplus}(x,t) &= \left\langle \sum_{n=1}^{L-x}\frac{1}{\alpha^{n+x}}\spin_{n+x} (t) \cdot \frac{1}{\alpha^{n}}\spin_n (0)\right\rangle\\
    &=  \sum_{n=1}^{L-x} \left\langle\frac{1}{\alpha^{n+x}}\spin_{n+x} (0) \cdot \frac{1}{\alpha^{n}}\spin_n (-t)\right\rangle \hspace{1cm}\text{(using time translational symmetry)}\\
    &=   \sum_{n'=x+1}^{L}\left\langle\frac{1}{\alpha^{n'}}\spin_{n'} (0) \cdot \frac{1}{\alpha^{n'-x}}\spin_{n'-x}(-t)\right\rangle \hspace{1cm} \text{(substitute $n' = n+x$)}\\
    &=  \sum_{n'= -(-x) }^{L-1}\left\langle \frac{1}{\alpha^{n'+(-x)}}\spin_{n'+(-x)}(-t)\cdot\frac{1}{\alpha^{n'}}\spin_{n'} (0)  \right\rangle \hspace{1cm} \text{(and $(-x) < 0$)}\\
    \implies C_{\magplus}(x,t) &= C_{\magplus}(-x,-t)
    \end{split}
\end{equation}
Hence, $C_{\magplus}(x,t) = C_{\magplus}(x,-t) = C_{\magplus}(-x,-t)$. Thus, it is symmetrical about $ x= 0$.

The spread of $C_{\mathcal{E}}(x,t)$ is symmetric about $x =0$. The proof will be similar to that of the previous case of $C_{\mathcal{M}}(x,t)$. First, we will prove that $C_{\mathcal{E}}(x,t) = C_{\mathcal{E}}(x,-t)$. If we invert the time i.e. $t \to -t$ and also make the substitution $\bm{\tilde{S}_n(t)} = -\spin_n(-t)$, then,
$\frac{d\bm{\tilde{S}}_n(t)}{dt} = \bm{\tilde{S}}_n(t) \times \left(\bm{\tilde{S}}_{n+1}(t) + \alpha \bm{\tilde{S}}_{n-1}(t)\right)$. These substitutions leave the form of \eqref{eqn:hybrid_dynmk1} invariant. Hence,
        \begin{equation}
    \begin{split}
        C_{\mathcal{E}}(x,t) &= \mathop{\sum\nolimits'}_n \left\langle\frac{1}{\alpha^{n+x+1}}(\bm{\tilde{S}}_{n+x} (t) \cdot \bm{\tilde{S}}_{n+x+1} (t))\frac{1}{\alpha^{n+1}}(\bm{\tilde{S}}_{n} (0) \cdot \bm{\tilde{S}}_{n+1} (0))\right\rangle\\
        &=\mathop{\sum\nolimits'}_n \left\langle \frac{1}{\alpha^{n+x+1}}((-\spin_{n+r} (-t))\cdot (-\spin_{n+x+1} (-t)))\frac{1}{\alpha^{n+1}}((-\spin_{n} (0)) \cdot (-\spin_{n+1} (0)))\right\rangle\\
         &= \mathop{\sum\nolimits'}_n \left\langle\frac{1}{\alpha^{n+x+1}}(\spin_{n+x} (-t) \cdot \spin_{n+x+1} (-t))\frac{1}{\alpha^{n+1}}(\spin_{n} (0) \cdot \spin_{n+1} (0))\right\rangle\\
    \implies C_{\mathcal{E}}(x,t) &= C_{\mathcal{E}}(x,-t)
    \end{split}
    \end{equation}
    Now, we will show $C_{\mathcal{E}}(x,t) = C_{\mathcal{E}}(-x,-t)$.
    Without loss of generality, consider $x > 0$. Then,
    \begin{equation}
    \begin{split}
    C_{\mathcal{E}}(x,t) &= \left\langle \sum_{n=1}^{L-x-1}\frac{1}{\alpha^{n+x+1}}(\spin_{n+x} (t) \cdot \spin_{n+x+1} (t))\frac{1}{\alpha^{n+1}}(\spin_{n} (0) \cdot \spin_{n+1} (0))\right\rangle\\
    &=  \sum_{n=1}^{L-x-1}\left\langle\frac{1}{\alpha^{n+x+1}}(\spin_{n+x} (0) \cdot \spin_{n+r+1} (0))\frac{1}{\alpha^{n+1}}(\spin_{n} (-t) \cdot \spin_{n+1} (-t))\right\rangle\\
    &=\sum_{n'=1}^{L-1}\left\langle\frac{1}{\alpha^{n'+1}}(\spin_{n'} (0) \cdot \spin_{n'+1} (0))\frac{1}{\alpha^{n'-x}}(\spin_{n'-x} (-t) \cdot \spin_{n'-x+1} (-t))\right\rangle \hspace{0.5cm} \text{($n' = n+x$)}\\
    &=  \sum_{n'=-(-x)+1}^{L-1}\left\langle\frac{1}{\alpha^{n'+1}}(\spin_{n'} (0) \cdot \spin_{n'+1} (0))\frac{1}{\alpha^{n'-x}}(\spin_{n'+(-x)} (-t) \cdot \spin_{n'+(-x)+1} (-t))\right\rangle \\
    \implies C_{\mathcal{E}}(x,t) &= C_{\mathcal{E}}(-x,-t)
    \end{split}
\end{equation}
 Hence, $C_{\mathcal{E}}(x,t) = C_{\mathcal{E}}(x,-t) = C_{\mathcal{E}}(-x,-t)$. Thus, it is symmetrical about $x= 0$.

\subsection{{Conservation of area under the curve for conserved-field correlators in the generic nonreciprocal model}}\label{App:Conserved Area_A_E}

{Using \eqref{pseudoenergycorr}, the sum over sites of the pseudo-energy correlator 
$C_{\mathcal{E}}(x,t)$
can be simplified:}
\begin{equation}
    \begin{split}
        A_{\mathcal{E}}(t)  &= \sum_{x = -(L-2)}^{L-2} C_{\mathcal{E}}(x,t)\\
        &= \sum_{x = -(L-2)}^{L-2} \left\langle \sum_n'\frac{1}{\alpha^{n+x+1}}(\spin_{n+x} (t) \cdot \spin_{n+x+1} (t))\frac{1}{\alpha^{n+1}}(\spin_{n} (0) \cdot \spin_{n+1} (0))\right\rangle\\
        &=\left\langle \left(\sum_{n'=1}^{L-1}\frac{1}{\alpha^{n'+1}}\spin_{n'} (t)\cdot \spin_{n' +1}(t)\right)\left(\sum_{n=1}^{L-1}\frac{1}{\alpha^{n+1}}\spin_n (0)\cdot \spin_{n+1} (0)\right)\right\rangle
    \end{split}
\end{equation}
The first term in the parentheses is the conserved quantity, and it is the only term that has a time index to it. Thus time derivative of  $A_{\mathcal{E}}$ vanishes.
\begin{equation}
    \frac{dA_{\mathcal{E}}}{dt} = \left\langle \left(\frac{d}{dt}\sum_{n'=1}^{L-1}\frac{1}{\alpha^{n'+1}}\spin_{n'} (t)\cdot \spin_{n' +1}(t)\right)\left(\sum_{n=1}^{L-1}\frac{1}{\alpha^{n+1}}\spin_n (0)\cdot \spin_{n+1} (0)\right)\right\rangle = 0\,. 
\label{eqn:ConservedArea_AE}
\end{equation}

\section{Comparison with Heisenberg dynamics}
\label{App:Hsbg_comparison}
\begin{figure*}[htp]
\centering
\begin{subfigure}{0.49\textwidth}
  \centering
  \includegraphics[width=\linewidth]{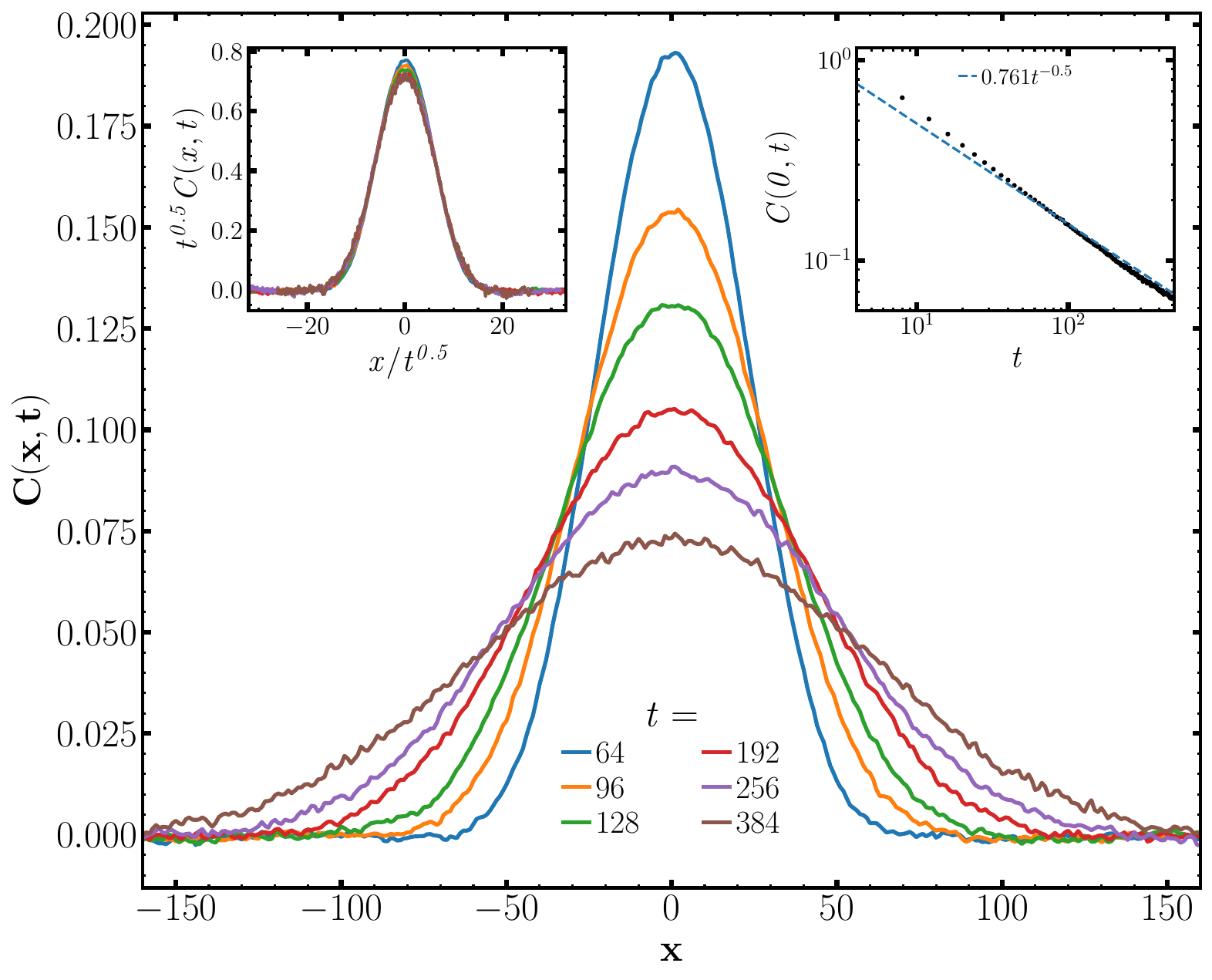}
  \caption{$C_{M}(x,t)$}
  \label{fig:Mxt_correlations}
\end{subfigure}\hfill
\begin{subfigure}{0.49\textwidth}
  \centering
  \includegraphics[width=\linewidth]{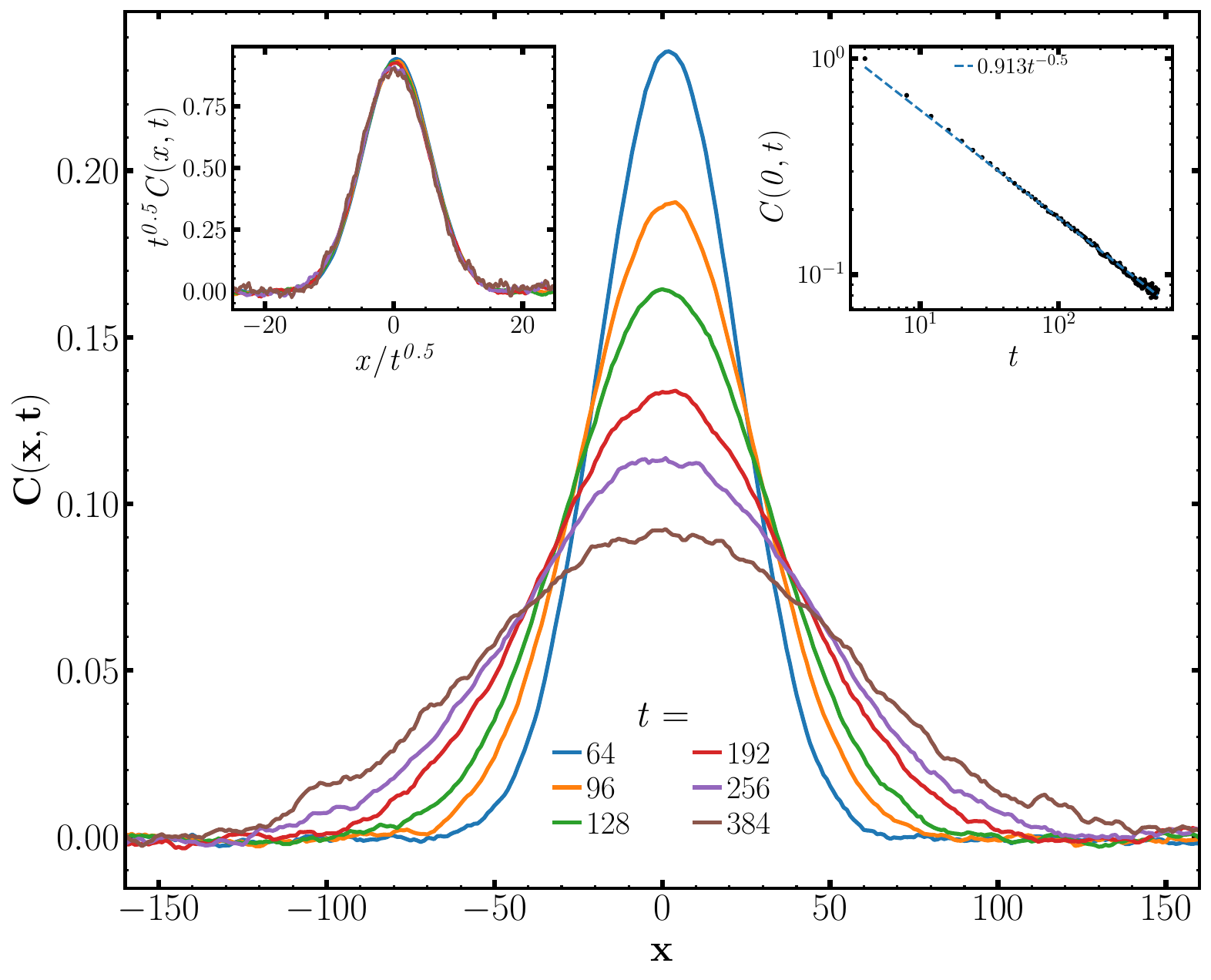}
  \caption{$C_{E}(x,t)$}
  \label{fig:CExt_correlations}
\end{subfigure}
\caption{ \small Two-point correlation functions of: (a) the  magnetization, $C_{M}(x,t)$ and (b) the energy , $C_{E}(x,t)$, with $x \in \{1,L\}$ for $L=512$, sampled over 5000 initial configurations as functions of $x$ for different values of $t$. The left inset of both panels shows the scaling collapse to a form $C(x,t) = t^{-1/2}f(x/t^{1/2})$ consistent with diffusion while the right shows a plot of $C(0,t)$ versus $t$ with a fit to $t^{-1/2}$.}
\label{fig:MM_E_corrxt}
\end{figure*}

\begin{figure*}[htp]
\centering
\begin{subfigure}{0.49\textwidth}
  \centering
  \includegraphics[width=\linewidth]{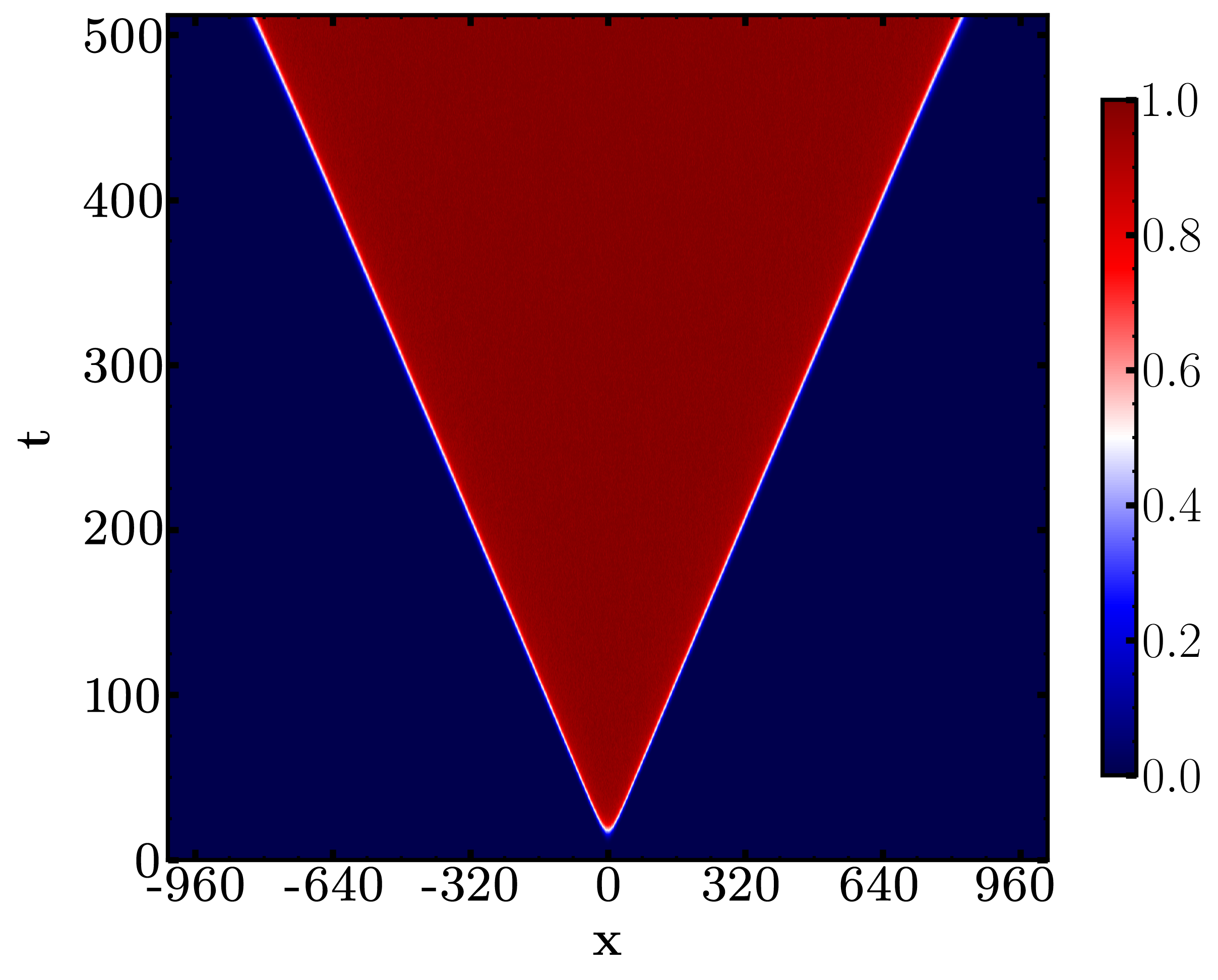}
  \caption{}
  \label{fig:Dxt_hsbg}
\end{subfigure}\hfill
\begin{subfigure}{0.49\textwidth}
  \centering
  \includegraphics[width=\linewidth]{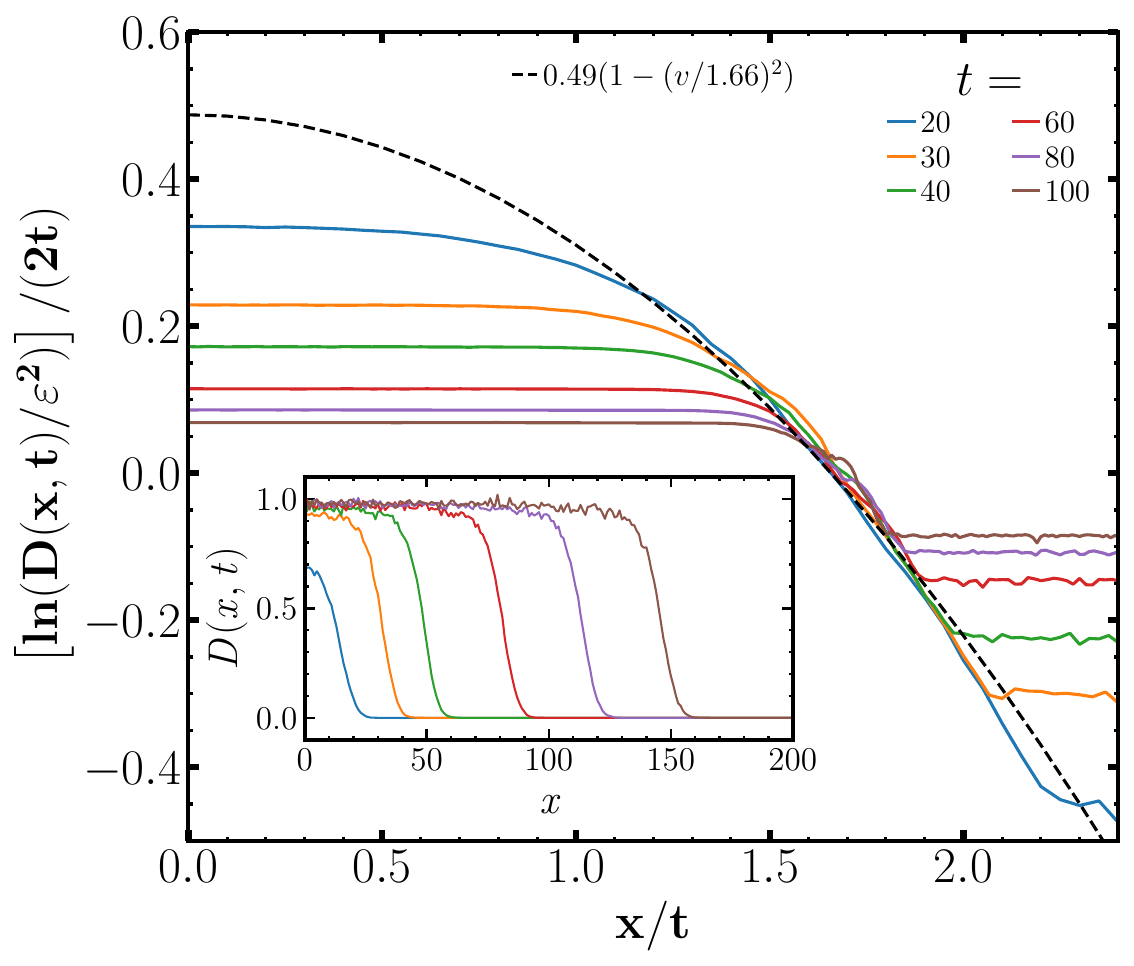}
  \caption{}
  \label{fig:logDxt_hsbg}
\end{subfigure}
\caption{\small \textit{(Left)} Colourmap of the decorrelator $D(x,t)$ calculated by averaging over pairs of initial conditions that differ only in the value of the spin ${\bf S}$ at the site $x=0$. It can be seen that this initial disturbance spreads ballistically from which the butterfly velocity $v_B=1.66$ can be obtained. \textit{(Right)} The decorrelator given by the expression $\log(D(x,t)/\varepsilon^2) = 2 \kappa t (1 - (x/v_Bt)^2)$ plotted as a function of $x/t$ for different values of $t$. As expected, it can be seen that there is a collapse of the curves in the vicinity of the front from which the Lyapunov exponent $\approx 0.49$ can be extracted. The inset shows $D(x,t)$ as a function of $x$ for different values of $t$. The existence of a front can be seen from the rapid decrease in the value of $D(x,t)$ as a function of $t$ (inset).}
\label{fig:Dxt_logDxt_hsbg}
\end{figure*}

\renewcommand{\theequation}{D.\arabic{equation}}
\setcounter{equation}{0}
To test the accuracy of our numerical code, we first ran the simulation for the reciprocal Heisenberg spin dynamics and calculated the conserved quantities -- standard magnetization and energy density. We confirmed that the conserved quantities show diffusive behaviour (Fig.\ref{fig:MM_E_corrxt}). 
We also calculated the numerical values for the butterfly velocity and Lyapunov exponent for this case, and found them to be within expected error range of our simulation parameters, $v_B = 1.66(\pm 0.02) , \kappa = 0.49 (\pm 0.02)$ (Fig. \ref{fig:Dxt_logDxt_hsbg}), with $\varepsilon = 10^{-3}, L = 2048, \Delta t = 0.001-0.002$.  
\comnt{
Finally, we also find the power-law associated with the broadening of the decorrelator front (Fig.\ref{fig:Arrivaltime_sup}). The approach here involves comparing the Decorrelator to a threshold value, $D_{0} = 100 \varepsilon^2$ and mark the time at which the single configuration arrival-front exceeds this threshold value, $D(x,t) \geq D_{0}$. 
Collecting this data for several samples ($\sim 10^4$) we see that the arrival-front of the decorrelator broadens with time. The distribution of the arrival-times from the mean arrival-time (the slope of which with respect to the position is just the inverse arrival velocity), shows a collapse when fit to a $1/3$ power-law.
}

\newpage
\bibliography{references}
\end{document}